\documentstyle[aps, preprint,eqsecnum, psfrag, graphicx, verbatim]{revtex}

\tightenlines

\begin{document}

\draft

\title{Renormalization group for network models of Quantum Hall transitions}
\author{ { Denis Bernard$^\spadesuit$}\footnote{Member of the
C. N. R. S.}   and  Andr\'e  LeClair$^{\clubsuit}$}
\address{$^\spadesuit$ Service de Physique Th\'eorique de Saclay
\footnote{Laboratoire de la Direction des Sciences de la Mati\`ere 
du Commisariat \`a l'Energie Atomique, URA 2306 du C. N. R. S.}
F-91191, Gif-sur-Yvette, France. }  
\address{$^\clubsuit$ Newman Laboratory, Cornell University, Ithaca, NY
14853.}
\date{July 2001}
\maketitle

\begin{abstract}

We analyze in detail the renormalization group flows which follow from
the recently proposed all orders $\beta$ functions for the Chalker-Coddington
network model.  The flows in the physical regime reach a true singularity 
after a finite scale transformation.  Other flows are regular and we identify
the asymptotic directions. One direction is in the same universality
class as the disordered $XY$ model. 
 The all orders $\beta$ function is computed for
the network model of the spin Quantum Hall transition and the flows are shown
to have similar properties.  It is argued that fixed points of general current-current
interactions in 2d should correspond to solutions of the Virasoro master equation. 
Based on this we identify two coset conformal field theories $osp(2N|2N)_1 /u(1)_0$
and $osp(4N|4N)_1/su(2)_0$  as possible fixed points and study the resulting
multifractal properties.  
We also obtain a scaling relation between the typical amplitude 
exponent $\alpha_0$ and the typical point contact conductance exponent
$X_t$ which is expected to hold when the density  
of states is constant.

\end{abstract}

\vskip 0.2cm

%
%

\def\dXY{dXY} 
\def\o{{8}}
\def\go{{g_\o}}
\def\debut{ \begin{eqnarray} }
\def\fin{ \end{eqnarray} }
\def\non{ \nonumber }
%
%
\def\betaf{$\beta~$}
\def\oti{{\otimes}}
\def\bra#1{{\langle #1 |  }}
\def\lb{ \left[ }
\def\rb{ \right]  }
\def\tilde{\widetilde}
\def\bar{\overline}
\def\hat{\widehat}
\def\*{\star}
\def\[{\left[}
\def\]{\right]}
\def\({\left(}          \def\BL{\Bigr(}
\def\){\right)}         \def\BR{\Bigr)}
        \def\BBL{\lb}
        \def\BBR{\rb}
%
%
\def\gp{g_+}
\def\betab{{\bar{\beta}}}
\def\zb{{\bar{z} }}
\def\zbar{{\bar{z} }}
\def\frac#1#2{{#1 \over #2}}
\def\inv#1{{1 \over #1}}
\def\half{{1 \over 2}}
\def\d{\partial}
\def\der#1{{\partial \over \partial #1}}
\def\dd#1#2{{\partial #1 \over \partial #2}}
\def\vev#1{\langle #1 \rangle}
\def\ket#1{ | #1 \rangle}
\def\rvac{\hbox{$\vert 0\rangle$}}
\def\lvac{\hbox{$\langle 0 \vert $}}
\def\2pi{\hbox{$2\pi i$}}
\def\e#1{{\rm e}^{^{\textstyle #1}}}
\def\grad#1{\,\nabla\!_{{#1}}\,}
\def\dsl{\raise.15ex\hbox{/}\kern-.57em\partial}
\def\Dsl{\,\raise.15ex\hbox{/}\mkern-.13.5mu D}
%
%
\def\dx{\frac{d^2 x}{2\pi}}
\def\th{\theta}         \def\Th{\Theta}
\def\ga{\gamma}         \def\Ga{\Gamma}
\def\be{\beta}
\def\al{\alpha}
\def\ep{\epsilon}
\def\vep{\varepsilon}
\def\la{\lambda}        \def\La{\Lambda}
\def\de{\delta}         \def\De{\Delta}
\def\om{\omega}         \def\Om{\Omega}
\def\sig{\sigma}        \def\Sig{\Sigma}
\def\vphi{\varphi}
%
%
\def\CA{{\cal A}}       \def\CB{{\cal B}}       \def\CC{{\cal C}}
\def\CD{{\cal D}}       \def\CE{{\cal E}}       \def\CF{{\cal F}}
\def\CG{{\cal G}}       \def\CH{{\cal H}}       \def\CI{{\cal J}}
\def\CJ{{\cal J}}       \def\CK{{\cal K}}       \def\CL{{\cal L}}
\def\CM{{\cal M}}       \def\CN{{\cal N}}       \def\CO{{\cal O}}
\def\CP{{\cal P}}       \def\CQ{{\cal Q}}       \def\CR{{\cal R}}
\def\CS{{\cal S}}       \def\CT{{\cal T}}       \def\CU{{\cal U}}
\def\CV{{\cal V}}       \def\CW{{\cal W}}       \def\CX{{\cal X}}
\def\CY{{\cal Y}}       \def\CZ{{\cal Z}}

\def\rvac{\hbox{$\vert 0\rangle$}}
\def\lvac{\hbox{$\langle 0 \vert $}}
\def\comm#1#2{ \BBL\ #1\ ,\ #2 \BBR }
\def\2pi{\hbox{$2\pi i$}}
\def\e#1{{\rm e}^{^{\textstyle #1}}}
\def\grad#1{\,\nabla\!_{{#1}}\,}
\def\dsl{\raise.15ex\hbox{/}\kern-.57em\partial}
\def\Dsl{\,\raise.15ex\hbox{/}\mkern-.13.5mu D}
%
%
%
\font\numbers=cmss12
\font\upright=cmu10 scaled\magstep1
\def\stroke{\vrule height8pt width0.4pt depth-0.1pt}
\def\topfleck{\vrule height8pt width0.5pt depth-5.9pt}
\def\botfleck{\vrule height2pt width0.5pt depth0.1pt}
\def\Zmath{\vcenter{\hbox{\numbers\rlap{\rlap{Z}\kern
0.8pt\topfleck}\kern 2.2pt
                   \rlap Z\kern 6pt\botfleck\kern 1pt}}}
\def\Qmath{\vcenter{\hbox{\upright\rlap{\rlap{Q}\kern
                   3.8pt\stroke}\phantom{Q}}}}
\def\Nmath{\vcenter{\hbox{\upright\rlap{I}\kern 1.7pt N}}}
\def\Cmath{\vcenter{\hbox{\upright\rlap{\rlap{C}\kern
                   3.8pt\stroke}\phantom{C}}}}
\def\Rmath{\vcenter{\hbox{\upright\rlap{I}\kern 1.7pt R}}}
\def\Z{\ifmmode\Zmath\else$\Zmath$\fi}
\def\Q{\ifmmode\Qmath\else$\Qmath$\fi}
\def\N{\ifmmode\Nmath\else$\Nmath$\fi}
\def\C{\ifmmode\Cmath\else$\Cmath$\fi}
\def\R{\ifmmode\Rmath\else$\Rmath$\fi}





\def\beq{\begin{equation}}
\def\eeq{\end{equation}}
\def\barray{\begin{eqnarray}}
\def\earray{\end{eqnarray}}
\def\sigvec{\vec{\sigma}}
\def\alvec{\vec{\alpha}}
\def\stwo{\sqrt{2}}
\def\psib{\bar{\psi}}
\def\bbar{\bar{\beta}}
\def\zb{{\bar{z}}}
\def\gal{g_\alpha}
\def\gm{g_-}
\def\gp{g_+}
\def\ga{g_a}
\def\gs{g_s}
\def\gc{g_c}
\def\go{g_\o} 
\def\Psib{\bar{\Psi}}
\def\Hb{\bar{H}}
\def\Jb{\bar{J}}
\def\Ct{\tilde{C}}

\section{Introduction}

The Chalker-Coddington network model is perhaps the simplest model believed
to be in the universality class of the Quantum Hall transition\cite{CC}.  
The model can be mapped to a free Dirac fermion theory with three independent
random potentials\cite{Ludwig}\cite{ChalkerCho}.  The renormalization group (RG)
fixed point of the disorder averaged effective action is expected to capture
the critical properties of the transition.  

Recently,  an all orders $\beta$ function was proposed for the network model\cite{Lec}.  
The main difficulty encountered in interpreting  the resulting RG flows is 
the occurance of singularities in the $\beta$ function.
For the case of anisotropic $su(2)$, by using duality and certain topological
identifications it was shown in \cite{BLdual} that all RG trajectories 
were regular in the sense that all flows could be extended to arbitrarily
large and small length scales without encountering any true singularities.  
Further checks of the $\beta$ function in the case of anisotropic $su(2)$
described in \cite{BLdual} provided additional support that the $\beta$ function
could be analytically continued beyond the domain of convergence of the
perturbation series.  

The main purpose of this work is to analyze in detail the RG flows of the 
Chalker-Coddington network model and also the network model for the spin Quantum
Hall transition\cite{sqhe,sqhe2,percolation} and to attempt to resolve the
possible singular flows using the ideas in \cite{BLdual}.    
There are large classes of flows that are regular, and for these we identify
two universal  attractive directions at large distances, referred to as 
$dXY$  and $gX$ below.  (Some of the phases listed as 
distinct in \cite{Lec}  become identified since we now understand
how to cross the singularities.) 
The $dXY$ phase is in the same universality class 
as a disordered $XY$ model or Gade-Wegner class\cite{Gade} 
and is equivalent to the model studied in \cite{GLL}.   
Unfortunately the flows in the physical domain of the network model with all couplings
(which are variances of disordered potentials) 
positive are pathological in the sense that they encounter a true singularity
after a finite scale transformation and the flow cannot be continued to larger
length scales.    These singular flows seem to be related to the fact that there
are level $0$ current algebras in these theories which arise naturally in
the supersymmetric method for disorder averaging.

Putting aside the $\beta$ function,  we argue that the possible fixed points
of general anisotropic left/right current-current interactions should correspond
to the solutions of the Virasoro master equation\cite{Halpern}
which are built upon the same
tensors that define the current interactions and thus preserve the global symmetries.
We classify these solutions for both network models and show that they correspond
to current-algebra cosets.  
Since this argument is independent of the
$\beta$ function,  it remains a possibility that a proper resolution of
the singular flows could lead to these fixed points.  
 It was argued in \cite{Lec}
that such cosets can arise as fixed points of the RG when the coupling
to a current subalgebra flows to infinity and the current subalgebra is gapped out.
 Some of the regular flows actually realize 
these fixed points,  in particular phase $C$ in the unphysical regime of 
spin network model flows to $osp(4|4)_1/osp(2|2)_{-2}$.  
By doubling the degrees of freedom as in \cite{GLL}, this may correspond to 
a physical model of interest.  For the spin network model one solution corresponds
to the coset $osp(4N|4N)_1/su(2)_0$, where $N$ is the number of copies,
and this fixed point gives some of the
same exponents as percolation\cite{percolation} though the equivalence of the
two theories is still a matter of debate.

For the $u(1)$ network model the only zeros of the $\beta$ function correspond
to a purely random $u(1)$ gauge field.  We point out that the stress tensor 
of this line of fixed points corresponding to the strength of the disorder 
does not satisfy the Virasoro algebra,  which apparently has not been noticed
before.  There is a special point on this line that leads to a constant density of
states,  and can formally be viewed as ``$osp(2N|2N)_1/u(1)_0$'' as far as the computation
of critical exponents is concerned.   (This is not a true coset since the
$u(1)_0$ does not commute with the stress tensor, however we find the coset
notation convenient and will use it in the sequel.)    This $c=0$ conformal field theory
is different from the $c=-2$ theory $osp(2N|2N)_1/u(1)\otimes u(1)$ considered
in \cite{Gurarie2}, nor is it the same as the theories considered in
\cite{Zirn,Tsvelik0}.  
 
We wish to emphasize that due to the singularities of the RG flow
in the physical
regime  of the network models we cannot show that the above cosets 
genuinely arise as fixed points. However, based on our analysis of
the Virasoro master equation,   they remain as interesting candidates 
for the correct fixed point and are worth further investigation, even
with the aim of ruling them out.   
In this paper we thus 
also explore the multifractal properties of the above coset theories,
and  show that these cosets lead to multifractality in a very natural way.  
This multifractality is an important constraint on the critical theory
and since it is not a generic feature of the kinds of fixed points that
can arise, it  can  serve as a criterion for ruling out some theories.   
In particular we show that the coset $osp(2N|2N)_1/u(1)\otimes u(1)$ does
not exhibit multifractality, whereas $osp(2N|2N)_1/u(1)_0$ does.  
For the coset $osp(4N|4N)_1 /su(2)_0$ we show that the typical
amplitude exponent $\alpha_0 = 9/4$, and the typical point contact conductance
exponent is $X_t = 1$.   Since the map to percolation\cite{percolation} 
does not apply to averages of higher moments of correlation functions, 
this result can serve as a useful test of the validity of the fixed
point $osp(4N|4N)_1/su(2)_0$.  
For the $u(1)_0$ coset, it turns out  $\alpha_0 =3$.   
However if one associates $\alpha_0$  with the twist fields    
which modify the boundary conditions 
then  $\alpha_0$ is again $9/4$ and this is close to 
numerical estimates
of $2.26$\cite{numerical2,numerical3}.  The small errors in the
most recent simulations however seem to rule out $9/4$\cite{numerical3}. 
We also obtain a scaling relation
between $\alpha_0$ and the typical point-contact conductance exponent $X_t$ 
which we expect to hold in a  theory with a constant density of states:
$X_t = 2(\alpha_0 - 2)$\cite{refer}.  

Our results are presented as follows.  In section II we extend the computation 
in \cite{Lec} to $N$ copies, and show, as expected, that the $\beta$ functions
are independent of $N$.   We present a remarkable strong-weak coupling duality of the
$\beta$ function.  We then describe all possible flows, and describe the nature
of the singular flows.  In section III we extend this analysis to the network 
model for the spin Quantum Hall transition and find similar properties.  
In section IV  we present our results on the multifractality.   The subsections
on the Virasoro master equation and multifractality are logically independent of
the sections describing the RG flows and can be read separately.

\section{The Chalker-Coddington network model}

The Chalker-Coddington network model\cite{CC} is expected to be in the same
universality class as the Dirac Hamiltonian with random mass,
potential and abelian gauge fields\cite{ChalkerCho}:
\debut
H = \left(\matrix{ V + m & -i \d_\zb + A_\zb 
         \cr -i \d_z + A_z & V - m \cr} \right) 
\label{HamilCC}
\fin
where $A,m, V$ are centered gaussian random potentials with variance
$g_a, g_m $ and $g_v$ respectively.  The physical regime of the network
model thus corresponds to all $g$ positive.  
The couplings $g_a, g_v , g_m$ correspond to randomness in the
link phases, the flux per plaquet and the tunneling at the nodes
respectively.   
This model also arose in the work\cite{Ludwig}.   The case
$g_m=g_v=0$ was studied in \cite{Ludwig,Mudry,Carg},
and the case $g_m + g_v = 0$ in \cite{GLL}. 
The one-loop \betaf function for all $g$'s non-zero was computed in
\cite{Carg}.  The all-orders \betaf function was proposed in 
\cite{Lec}.

\subsection{N-Copies effective action.} 

Following the conventions in \cite{Carg,Lec} we study the action
\barray
\label{2.1}
S &=&  \int \dx
\Bigl[  \psib_- (\d_z - i A_z ) \psib_+ + \psi_- (\d_\zbar
-i A_\zbar ) \psi_+
\\
&& ~~~~~~~~~~~~~-i V (\psib_- \psi_+  + \psi_- \psib_+ )
    ~ -im ( \psib_- \psi_+ - \psi_- \psib_+ ) \non
\Bigr]
\earray
where $z=x+iy$, $\zbar = x-iy$. 
The above action
corresponds to $S=i \int  \psi^\star H \psi$ where 
$H$ is the hamiltonian of the network model.  To study
the Green functions at energy $\CE$, one lets 
$H\to H-\CE$ which leads to a term in the action 
\beq
\label{phiE}
S_\CE = i \int \dx ~ \CE ~ \Phi_\CE , ~~~~~~~~
\Phi_\CE = \psib_- \psi_+ + \psi_- \psib_+ 
\eeq
The retarded and advanced Green functions then correspond to 
$\CE = E \pm i \epsilon$ with $\epsilon$ small and positive.

To study the delocalization transition one is interested in the
critical points of the disorder averaged theory at $\CE =0$.  Since
$\Phi_\CE$ corresponds to various mass terms, the theory can not be  
critical when $\CE \not=0$.  
The \betaf function we determine below is for $\CE=0$, and for
this reason it is independent of whether the various copies
are retarded or advanced. One can in fact convert advanced copies
into retarded ones by flipping the sign of the right-moving
fields $\psib_\pm \to - \psib_\pm$.  Since this does not change the 
$\psib_\pm$ operator products, the \betaf function is unchanged.
Once one flows to a critical point,
one needs to reintroduce $\CE$ in order for instance to compute
conductivities.   This can be compared with the sigma model approach.
The latter approach was applied to the  coupling sub-manifold
$g_m + g_v = 0$ 
of the network model in \cite{GLL}.   An important point is
that the saddle point approximation that leads to a sigma model
occurs in the $\CE=0$ theory.  Thus one expects that  
our RG flows should in principle capture  these saddle points. 

In order to study disorder averages of mixed products of retarded
and advanced Green functions we introduce $N$ copies of the action
(\ref{2.1}), with fields
$\psi_\pm^\alpha$, $\psib_\pm^\alpha$ where $\alpha = 1,..,N$ is a
copy, or flavor,  index.  
The disorder averaging is then performed using the supersymmetric
method by introducing ghosts $\beta_\pm^\alpha, \betab_\pm^\alpha$. 
To present the effective action in a compact form, we introduce
the $2N$ component vectors $\Psi_\pm^a$, $a=1,..,2N$, and we let
$\Psi_\pm = ( \psi^1_\pm ,...,\psi^N_\pm,
\beta^1_\pm,..., \beta^N_\pm)$
and similarly for $\Psib_\pm$.  
We define a grade $[a]=0,1$ where $[a] = 1$ for the fermionic
components $\psi$ and $[a]=0$ for the bosonic components $\beta$. 
The effective (disorder averaged) action is then 
\beq
\label{2.4} 
S = S_{\rm free} + \int \dx \( g_+ \CO^+ + g_- \CO^- + g_a \CO^a \) 
\eeq
where $g_\pm = g_v \pm g_m$, and  $S_{\rm free}$ the free $c=0$ 
conformal field theory of the $\psi, \beta$ fields, 
\beq
S_{\rm free} = \int \dx \( \Psi_- \d_\zbar \Psi_+ + \Psib_- \d_z\Psib_+ \) 
\label{Sfree}
\eeq
where $\Psib_- \Psi_+ = \sum_a \Psib_-^a \Psi_+^a$, etc.
The operator product expansion (OPE) implied by the free action
(\ref{Sfree}) is 
\beq
\label{2.6}
\Psi^a_+ (z) \Psi^b_- (0) \sim \inv{z} \delta^{ab} ,~~~~~
\Psi^a_- (z) \Psi^b_+ (0) \sim - \frac{(-)^{[a]}}{z}  \delta^{ab} ,
\eeq
and similarly for $\Psib$.  
In order to compute the \betaf function we express the perturbing
operators in eq. (\ref{2.4}) as left-right current-current
interactions.   Define the left-moving currents
\beq
\label{2.7}
J^{ab}_\pm = \Psi^a_\pm \Psi^b_\pm  
\quad {\rm and} \quad
H^{ab} = \Psi^a_+ \Psi^b_- ,
\eeq
and similarly for the right movers $\Jb^{ab}_\pm$ and $ \Hb^{ab}$. 
The perturbing operators can then be written as 
\barray
\nonumber
\CO^+ &=& \inv{2} \( (\Psib_- \Psi_+ )^2 + (\Psi_- \Psib_+ )^2 \)
=\inv{2} \sum_{a,b=1}^{2N} (-)^{[a]} 
\(  J^{ab}_+ \Jb^{ba}_- + J^{ab}_- \Jb^{ba}_+ \) \\
\label{2.8} 
\CO^- &=&  (\Psib_- \Psi_+)(\Psi_- \Psib_+ )
=\sum_{a,b=1}^{2N} (-)^{[a]} H^{ab} \Hb^{ba} \\
\nonumber
\CO^a &=&  (\Psi_- \Psi_+ )(\Psib_- \Psib_+ ) 
=\sum_{a,b=1}^{2N} (-)^{[a]+[b]} H^{aa} \Hb^{bb} 
\earray
The currents (\ref{2.7}) satisfy the $osp(2N|2N)_k$ current
algebra with level $k=1$.  

\subsection{$\beta$eta functions and duality.}

We now  extend the computation in \cite{Lec} 
to $N$-copies by using the previous compact notation.  
The resulting \betaf function is independent of $N$. 
 We then describe the  strong-weak coupling duality of the
\betaf function that is important for
extending  the flows to all scales.  

The effective action can be expressed as 
\beq
\label{action}
S = S_{\rm wzw}^{\CG_k} + 
\int \dx \sum_A g_A \CO^A  ~~~~~{\rm with} ~~  \CO^A = d^A_{ab} J^a \Jb^b
\eeq
where $S_{\rm wzw}^{\CG_k}$  is the action for the conformal WZW model 
with current algebra symmetry $\CG_k$, where $k$ is the level, and $J^a$ 
are the $\CG_k$ currents.  
For the network model $\CG_k = osp(2N|2N)_1$, and the 
 tensors $d^A_{ab}$ are implicitly defined by eqs.(\ref{2.8}).
The \betaf function proposed in \cite{GLM} is expressed in terms
of some OPE coefficients $C, D, \Ct$.
Let $T^A$ be the left-moving operator, 
\beq
\label{3.5b}
T^A (z) = d^A_{ab} J^a (z) J^b (z)
\eeq
Then the RG data can be computed from the OPE's 
\barray
\label{2.10}
\CO^A (z, \zbar) \CO^B (0) &\sim& \inv{z\zbar}
C^{AB}_C \> \CO^C (0)\\ \nonumber
T^A (z) \CO^B (0) &\sim&  \inv{z^2}
\( 2k D^{AB}_C +  \Ct^{AB}_C  \) \CO^C (0)
\earray

Specializing to the $N$-copy network model one finds that
$C,D,\Ct$ are independent of $N$ due to the fact that 
$\sum_{a=1}^{2N} (-)^{[a]} = 0$, which is equivalent to the statement
that the superdimension of $osp(2N|2N)$ is zero.  
As expected, the \betaf function is then identical 
to the $N=1$ result in \cite{Lec}:  
\barray
\nonumber
\beta_{g_+} &=&
\frac{8g_+\(2g_a(g_+^2+4)+(2-g_-)(g_+^2+2g_-)\)}{(4-g_+^2 )(2-g_-)^2 }\\
\beta_{g_-} &=& \frac{ 8 g_+^2 (2+g_-)^2 }{(4-g_+^2 )^2 }
\label{2.11}\\ 
\nonumber
\beta_{g_a} &=&  \frac{ 4 \( (g_+^2 - g_-^2)(16-g_+^2 g_-^2 )
+ 4 \ga \gp^2 (2+\gm) (2-\gm)^2 \) }{ (4-\gp^2)^2 (2-\gm)^2 }
\earray
Here $\beta = dg/d\tau$ where $\tau = \log r$ is the RG `time'
and $r$ is a length scale. 
The only zero of the $\beta$ function is at $g_\pm = 0$ with
$g_a$ arbitrary. 
We remark that $\beta_{g_-}$ is always positive for real couplings so
that $g_-$ is always increased by RG transformations.

The strong-weak coupling duality of the \betaf function is
of the following kind.  Let $g^*$ denote dual couplings
which are functions of $g$.  Suppose
\beq
\label{2.12}
\beta^* (g^*) = \frac{\d g^*}{\d g} \beta (g) = - \beta(g\to g^*)
\eeq
The above relation implies that if 
$g(r)$ is a solution of the RG equations, then so is
$g^*(r_0/r)$ for some $r_0$.   Furthermore, if  
$g$ is self-dual at some
scale $r_0$, i.e. $g=g^*$, then the ultra-violet (UV) and 
infra-red (IR) values of $g$ are related by duality:
$g_{IR} = g^*_{UV}$.

The \betaf function (\ref{2.11}) remarkably has such a duality with 
\beq
\label{2.14} 
g_+^* = \frac{4}{g_+} , ~~~~~g_-^* = \frac{4}{g_-} ,~~~~~
g_a^* = -\frac{4 g_a }{g_-^2}
\eeq
The self-dual points $g=g^*$ are $(g_+, g_-, g_a) 
= (\pm 2, \pm 2, 0)$.  

The above duality is one ingredient we will use to resolve
some difficulties encountered in \cite{Lec} in completely
interpreting the flows, 
as was done for anisotropic $su(2)$ in \cite{BLdual}. 
But as we shall see not all RG flows may be resolved in this way.
Also, unlike the $su(2)$ case, we have not found an RG invariant
for the above $\beta$ functions.







\subsection{Virasoro master equation.} 

The action (\ref{action}) has a well-defined classical stress tensor $T^{\rm class}_{\mu\nu} $.  
Introducing a 2d metric $g^{\mu\nu}$, then 
$T^{\rm class}_{\mu\nu} = - 4\pi {\delta S}/{\delta g^{\mu\nu}}$.  One finds that
this classical stress tensor is traceless, i.e. $g^{\mu\nu} T^{\rm class}_{\mu\nu} = 0$ 
and the left-moving conformal stress tensor $T = T_{zz}$ is 
\beq
\label{classic}
T^{\rm class} =  T^{\rm class}_{\rm wzw} - g_A T^A
\eeq
where the $T^A$ are defined in eq. (\ref{3.5b}),
and $T^{\rm class}_{\rm wzw}$ is the classical stress tensor for the WZW model which is
the $k\to \infty$ limit of the affine Sugawara stress tensor.  

In the quantum theory there are order $g^2$ corrections to the trace of the stress tensor 
and the conformal invariance is thus broken.  Based on the form (\ref{classic}) it is natural
to suppose that if the theory flows under RG to a conformally invariant fixed point, then
at the fixed point the stress-tensor 
 takes the form 
\beq
\label{fixed}
T_{\rm fixed ~ point} =  T_{\CG_k} -  \delta T ,  ~~~~~~~~{\rm with} 
~~~\delta T = h_A T^A
\eeq
where $T_{\CG_k}$ is the affine Sugawara stress-tensor and 
$h_A $ are constants which are unknown functions of the fixed point values of $g_A$.

The condition that a  stress tensor  $T$ satisfies the Virasoro algebra, 
i.e. 
\beq
\label{Tope}
T(z) T(0) \sim  \frac{c/2}{z^4} + \frac{2}{z^2} T(0) + .... , 
\eeq
is the so-called Virasoro master equation studied extensively in  
\cite{Halpern}.  
It is known that solutions come in pairs (K-conjugation), 
such that the sum of the stress tensors for a given pair
is the affine-Sugawara stress tensor $T_{\CG_k}$.
Thus $\delta T$ (and consequently $T_{\rm fixed~point}$) 
must be a solution of the master equation, which reads
\debut
2k D^{AB}_Ch_Ah_B + C^{AB}_Ch_Ah_B + 2 \tilde C^{AB}_Ch_Ah_B =h_C
\label{master}
\fin
where the coefficients $D^{AB}_C$, $C^{AB}_C$ and $\tilde C^{AB}_C$ are the same
as those determining the $\beta$ function 
defined in eqs.(\ref{2.10}). 
Though it is intriguing that the same coefficients $C, \tilde{C}, D$ 
determine both the $\beta$ function and the Virasoro master equation, 
we cannot show in complete generality that solutions of the master equation
correspond to zeros of the $\beta$ function.

Using the $C, \tilde{C}, D$
computed in Ref. \cite{Lec} one finds 2 pairs of solutions
to eq. (\ref{master}).   The fact that the solutions have
a simple interpretation serves as a check of the coefficients
$C, \tilde{C}, D$.  The first pair of solutions is
\beq
\label{pair1}
T_{osp(2N|2N)_1} = \inv{2} \( T^- - T^+ \) , ~~~~~~~~
T_0 = 0
\eeq
Here $T_{osp(2N|2N)_1}$ is the affine-Sugawara stress tensor
for the full current algebra and $T_0$ for the empty coset
$osp(2N|2N)_1 / osp(2N|2N)_1$.   The other pair of solutions
is
\beq
\label{pair2}
T_{gl(N|N)_1} = \inv{2} \( T^a - T^- \) , ~~~~~~~~~
T_{osp(2N|2N)_1 /gl(N|N)_1 } = T^- - \inv{2} T^a - \inv{2} T^+
\eeq
corresponding to the sub-current algebra $gl(N|N)_1$ of
$osp(2N|2N)_1$ and the indicated coset. 
($T_{\CG/\CH} = T_\CG - T_\CH$).  The two terms in $T_{gl(N|N)_1}$ 
correspond to the two independent quadratic Casimirs of 
$gl(N|N)$\cite{Rozansky}.  

One can form a level-1 representation of both $osp(2N|2N)$ and
$gl(N|N)$ with the same field content,  thus both $T_{osp(2N|2N)_1}$
and $T_{gl(N|N)_1}$ are equivalent to the free-field stress tensor
$T_{\rm free}$ 
of the $\psi, \beta$ fields.   This leads to the identity
\beq
\label{identity}
T_{\rm free} = T_{gl(N|N)_1} = T_{osp(2N|2N)_1} , 
~~~~\Rightarrow  T^- -\inv{2} T^a - \inv{2} T^+ = 0 
\eeq
This  is equivalent to the statement $T_{osp(2N|2N)_1/gl(N|N)_1} = 0$.  

It is important to note that the purely random $u(1)$  with only $g_a$ not equal
to zero does not appear as a solution of the master equation and is thus not a 
consistent stress-energy tensor.   
In terms of currents, 
\beq
\label{2.22}
T_{u(1)_0} \equiv  \inv{2} T^a = \inv{2} 
H^2 ,  ~~~~~~H \equiv \sum_{a=1}^{2N} (-)^{[a]} H^{aa} = H_\beta - H_\psi
\eeq
where
\beq
\label{2.23}
H_\beta = \sum_{\alpha=1}^N \beta_+^\alpha \beta_-^\alpha,
~~~~~H_\psi = \sum_{\alpha =1}^N \psi_+^\alpha \psi_-^\alpha
\eeq
These currents satisfy two $u(1)$ subalgebras:
\beq
\label{2.24}
H_\beta (z) H_\beta (0) \sim - H_\psi (z) H_\psi (0)
\sim - \frac{N}{z^2}
\eeq
Note that  $H(z) H(0) \sim {\rm reg.}$, i.e. since $H$ satisfies
a $u(1)$ current algebra at level-0, and $T_{u(1)_0} $ is formally
the Sugawara form for $u(1)_0$. 
The level being zero is what prevents $T_{u(1)_0}$ 
from being a good stress-tensor.  

Consider the class of stress-tensors
\beq
\label{Th}
T_h =  T_{\rm free}  - h T_{u(1)_0} = -\inv{2} T^- + (1-h) T_{u(1)_0} 
\eeq
The stress-tensor for the pure random $u(1)$ is of this form with $h=2g_a$,
and corresponds to the only zero of the $\beta$ function.  
This stress tensor satisfies:
\beq
\label{Thope}
T_h (z) T_h (0) \sim \frac{2}{z^2} \( T_h - h T_{u(1)_0} \) , ~~~
T_h (z) T_{u(1)_0}  (0) \sim \frac{2}{z^2}  T_{u(1)_0}  (0) , ~~~
T_{u(1)_0}  (z) T_{u(1)_0}  (0) \sim {\rm reg.} 
\eeq
The Virasoro algebra is spoiled by the $h$ term in the first equation.  
We should thus include this $c=0$ generalization of Virasoro as a possible fixed point.  

As far as the computation of conformal scaling dimensions is concerned we
can formally  view  the stress tensors (\ref{Th}) as  the coset  
"$osp(2N|2N)_1/u(1)_0$". 
However it is not a true coset since the $u(1)_0$ does not commute with
$T_h$.  
The point $h=1$ is special, in that $T_{u(1)_0}$ is precisely canceled
and $T_{h=1}$ reduces to $-T^-/2$, and this leads to a constant
density of states.  (See section IV.) 
Since the notation is convenient and descriptive we will refer to the
$T_{h=1}$ theory as $osp(2N|2N)_1/u(1)_0$. 
We emphasize that the latter theory 
 is not the same as the $c=-2$ theories $osp(2N|2N)_1/u(1)\otimes u(1)$ 
considered in \cite{Zirn,Gurarie2,GLL,Lec}  
where the $u(1)$'s are generated by $H_\psi$ and  $H_\beta$.

\subsection{Singular  RG trajectories in the Chalker-Coddington 
network model.}

There exists a small category of  RG trajectories which cannot
be extended consistently to arbitrarily large  RG  time.  
This  is probably linked
to the level zero current algebras appearing in these disordered systems.

These pathologies concern trajectories, or their duals, 
which enter the perturbative physical domain,
the latter  defined by requiring that the variances 
$g_v,\ g_m,\ g_a$ are all positive.
In a finite RG time these singular trajectories 
converge  to the singular
locii $g_\pm=2$ with $g_a>0$.

Let us first argue that the perturbative physical domain 
is stable under
renormalization until reaching the singularities at $g_\pm=\pm2$.
The boundaries of this domain are $g_a=0$, $g_+=2$ and $g_+\pm g_-=0$. 
 From eq.(\ref{2.11}), one verifies that $\beta_{g_a}>0$ for
$g_a>0$  and $0<|g_\pm|<2$. Hence, $g_a$ remains
positive as long as the RG trajectories are in this domain.
Similarly, one verifies that the scalar product 
$\beta_g\cdot n$, with $n$ the normals to the
boundaries pointing inwards, is positive on the boundaries.
Thus, RG flows which are initially inside the domain
cannot cross the boundaries and stay within it.

Since $g_-$ increases, the only direction in which
trajectories entering the physical domain from elsewhere in the phase space
 could go is towards the singular point $g_\pm=2$.
Furthermore, since in this domain $g_a$ is always increasing 
these physical trajectories reach $g_\pm=2$ with $g_a>0$.
This typical behavior may be checked by 
looking at the asymptotic behavior in the vicinity of the pole 
$(g_+ , g_- ) = (2+ \ep_+ , 2 + \ep_- )$.  Here the dominant terms
in the \betaf functions are 
\beq
\label{loops5bis}
\beta_{\ep_+} \simeq -\frac{64 g_a}{\ep_+ \ep_-^2} , \quad
\beta_{\ep_-} \simeq \frac{32}{\ep_+^2} , \quad
\beta_{g_a} \simeq \frac{16 g_a}{\ep_+^2}  
\eeq
Integrating these one finds
\beq
\label{loops6bis} 
g_a \simeq C\, \exp{\ep_-/2} , \quad 
\ep_+ \simeq C'\, \exp{2C/\ep_-} 
\eeq
with $C$ and $C'$ two integration constants.
Eq.(\ref{loops6bis}) shows that $g_a$ flows 
to a non-vanishing constant as $\ep_-\to 0$,
while $\ep_+$ vanishes exponentially. 
Since eq.(\ref{loops6bis}) involves two free parameters
these trajectories are generic enough in this domain.

Unfortunately, because these physical  flows reach 
the singularity at points distinct from the self-dual point
$(g_\pm=2,g_a=0)$, they cannot be extended smoothly beyond
this  point using duality and
possible additional identifications as explained in \cite{BLdual}.
Thus these RG trajectories are singular at a finite RG time,
ie. after a finite scale transformation. This pathological
behavior is in contradiction
to what is expected for a renormalizable field theory.
The singular flows are depicted in figure 3.  Trajectories can
enter the physical domain either from the $g_+=0$ plane or
from the point $g_\pm =\pm 2, g_a =0$.

\subsection{Regular RG trajectories in the Chalker-Coddington 
network model and phase diagram.}

Most of the flows are actually not singular. 
As was done for the anisotropic $su(2)$ model in ref.\cite{BLdual}, 
one may use duality arguments to
construct smooth extensions such that these flows may be extended
to arbitrarily  large scale. The main steps in resolving these flows are:
(i)  RG flows to the singular self-dual points are
extended beyond it using the duality and, 
(ii)  flows where the couplings blow up in finite RG time 
but do not cross self-dual points  
are extended with certain topological constructions with
the \betaf function serving as a consistent patching condition.
In this section we describe the typical
behaviors of these trajectories and identify  the resulting asymptotic
phases. As we will show, for this class of RG flows,
there are only two such directions referred to as
$\dXY$ and $gX$ below.

To begin describing these regular flows, consider the \betaf function when
$g_+ =0$. There, 
\beq
\label{dxy}
\beta_{g_-} = 0, ~~~~~~
 \beta_{g_a} = - {4g_-^2}/{(2-g_-)^2}
\eeq
Since $\beta_{g_-} = 0$, we can consider the flows as originating
or terminating at $g_+ = 0$. On the $g_+ =0$ plane,
the flows all originate from $g_a = +\infty$ since there $g_a$ always
decreases. Near $g_+ = 0$, one has
\beq
\label{qhe2}
\beta_{g_+} \simeq \frac{ 16g_+ (g_a - g_{a0})}{(2-g_-)^2} ,\quad
g_{a0} \equiv g_- (g_- -2)/4
\eeq
Thus, if $g_a > g_{a0}$ the flows are away from the $g_+=0$
plane, otherwise they are attracted to it. 
Since  $\beta_{g_a}<0$ and  $g_a$ decreases in this domain, 
trajectories which are attracted to the plane $g_+=0$ 
are more and more attracted to it, while those which flow away from it
may  flow back toward it if $g_a-g_{a0}$ is not large  enough.
This implies that some trajectories form loops, in the sense
that they begin and end at different points on the $g_+ = 0$ 
plane but with different values of $g_-$ and $g_a$.  (See Figure 1.)  

Near $g_+ = \infty$ one finds $\beta_{g_-} = 0$,  
$\beta_{g_a} = -{4g_-^2}/{(2-g_-)^2}$ and
\beq
\label{qhe3}
\beta_{g_+}\simeq - \frac{ 16g_+ (g_a - g_{a\infty})}{(2-g_-)^2} , \quad
g_{a\infty} \equiv  (g_- -2)/2
\eeq
Thus if $g_a > g_{a\infty}$, $g_+$ decreases away from $\infty$, otherwise
it continues to flow off to infinity. 
Since $\beta_{g_a}<0$, $g_a$ decreases there and trajectories
flowing off will escape more and more to infinity, while those with
$g_+$ decreasing may be repelled back to infinity if
$g_a-g_{a\infty}$ is not large enough, again forming
loops that are simply duals of the loops near $g_+=0$. 
(See Figure 1.)
Notice that $g_{a0}^*(g_-)=g_{a\infty}(g_-^*)$ in agreement with duality.

Flows that do not loop back to $g_+ = 0, \infty$ eventually
encounter the poles in the $\beta$ function in a finite RG time. 
Let us consider the ratios $\d g_+ / \d g_- = \beta_{g_+} / \beta_{g_-}$ 
and $\d g_+/\d g_a= \beta_{g_+} / \beta_{g_a}$.
They determine the direction of the flows.   
One finds that, except at the self dual points, both
$\d g_+ / \d g_-=0$ and $\d g_+ / \d g_a=0$
when $g_+$ approaches the singularities at $g_+ = \pm 2$.   
This implies that the flows near the poles
at $g_+ = \pm 2$ are tangent to the $g_+ = \pm 2$ planes
since $\beta_+\ll\beta_-\;,\beta_a$.  
Similarly $\d g_- / \d g_+ = 0$ and $\d g_- / \d g_a = 0$
when $g_- = \pm 2$. Thus  $\beta_-\ll\beta_+\;,\beta_a$
and the flows are tangent to the $g_- = \pm 2$ planes
near $g_- = \pm 2$. Consequently  if the flows 
indeed cross the singular planes, they must do so at the  
self-dual points of $g_\pm$.  In order for the flows
to continue smoothly through the poles one also needs $\d g_+ / \d g_-$ and
$\d g_+/\d g_a$ to be finite at the poles.  
Examining the latter one finds that this can occur  only when $g_a=0$.  
We thus conclude that if the flows indeed
cross the singular planes at finite angles, they must do so 
at the self-dual points $g=g^*$ for all $g$ simultaneously.  

The crossing of the singularities may be described more precisely.
Consider first the pole at $g_\pm=\pm 2$. Looking for flows
approaching the self-dual point $g_\pm=\pm 2,\ g_a=0$, one may
search for an asymptotic expansion of $2+g_-$ and $g_a$ as functions of
$\ep_+=g_+-2$. To lowest order, the equations for the trajectories are:
\debut
g_- &=& -2 - \ep_+ + 2Q_-\,\ep_+^4 - Q_- \ep_+^5 + \CO(\ep_+^6) 
\label{through1}\\
g_a &=& Q_-\,\ep_+^4 - Q_- \ep_+^5/2 + \CO(\ep_+^6) \non
\fin
with $Q_-$ a free parameter. These flows form a continuous
set of trajectories.  However since this is
only a one parameter family, they are not the most generic ones.
The $\beta$ function for $\ep_+$ is smooth,
$\beta_{g_+}=-2-\ep_++\cdots$ implying that $\ep_+$ decreases.
So these flows, coming from the domain $g_-<2$, $g_+>2$,
reach the self-dual point in a finite RG time and then cross  over.
(See Figure 2).

The flows which cross the upper self-dual point at $g_\pm=2$, $g_a=0$
either start with $g_+<2$ or $g_+>2$. In the first case, 
their trajectories are described by
\debut
g_-&=& 2+ \ep_+ + Q_+\ep_+^2 + Q_+^2 \ep_+^3 + \CO(\ep_+^4) 
\label{through2}\\
g_a&=& Q_+\ep_+^2/2 + Q_+^2 \ep_+^3/2 + \CO(\ep_+^4) \non
\fin
with $\ep_+=g_+-2$ and $Q_+$ a free parameter labeling them.
The $\beta$ function $\beta_{g_+}$ possesses a double pole,
$\beta_{g_+}=32(1 + (Q_+-2)\ep_+ +\cdots)/\ep_+^2$ so that 
$\ep_+^3= 96(\tau-\tau_*)+\cdots$. These flows reach the self-dual point
from $g_+<2$ in a finite RG time
but admit an extension beyond that point. As depicted in 
Figure 3 they then flow off to infinity.
In the second case, the trajectory equations are 
\debut 
g_-&=&2-\ep_++(4\tilde Q_++1)\ep_+^2/2+\cdots 
\label{through3}\\
g_a&=&\tilde Q_+\ep_+^2+\cdots \non
\fin
with $g_+=2+\ep_+$.
Again the $\beta$ function for $\ep_+$ has a double pole
$\beta_{g_+}=-32(1 + (4\tilde Q_++1)\ep_+ +\cdots)/\ep_+^2$ but
with opposite sign. Hence,  $\ep_+$ now decreases and,
after going through the self-dual point, these trajectories
flow towards the axis $g_+=0$. (See Figure 2.)

Let us now show  how one may resolve the blow-ups in finite RG time. 
This concerns flows toward $g_- , g_a = \infty$ that are dual
to those crossing the $g_+$ axis. 
Indeed, any trajectory that passes through
$g_- = 0$ is mapped by duality to two separate trajectories 
at $g_- = \pm \infty$. Their asymptotic at infinity are
\debut
\label{as2}
g_+ \to {\rm const.} , ~~~~~
g_a \simeq {\rm const'.}\,g_-^2 , ~~~~~
g_- \simeq \frac{(4-g_+^2)^2}{8g_+^2 (\tau_* - \tau)}
\fin
where $\tau_*$ is a constant.  
Eq.(\ref{as2}) shows that in a finite time $\tau_*$ the flow is to
$(g_+, g_- , g_a) \to ({\rm const. }, \infty, \infty)$.
We have to extend the flows to larger  scales.
Since the trajectories are continuous at $g_- = 0$,  
it is natural to demand that 
their dual be continuous at $g_- =\pm \infty$
so that flows towards $(g_- , g_a)= (\infty, \infty)$ 
continue at $(g_- , g_a)   = (-\infty, \infty)$.
Imposing this continuity requires identifying points at infinity
by changing the topology of the phase space in a way consistent
with the $\beta$ functions. To be more precise let us map points 
at infinity to finite distance by defining new coordinates 
$h_\pm,\,h_a$ by $h_+=g_+$, $h_-=1/g_-$ and $h_a=-g_a/g_-^2$.
We identify the points with coordinates $(h_+,h_a,h_-=0^+)$ and
$(h_+,h_a,h_-=0^-)$ with $h_a\not= 0$, or alternatively
\debut
\label{patch1}
(g_+ , g_- , g_a )\equiv  (g_+ , -g_- , g_a ), \quad {\rm as}\ 
|g_-|\to \infty,\ g_a/g_-^2\not= 0\ {\rm fixed} 
\fin
This is consistent with RG flows as the $\beta$ functions
$\beta_h = \beta_g\, (\d h/\d g)$ in the new coordinates 
are continuous when $h_-$ crosses the origin.
This proposal was described in greater detail for the $su(2)$ case 
in \cite{BLdual}. The identification (\ref{patch1}) changes the 
topology of the space of couplings. We can view this as gluing
multiple patches of coupling constant space along $g_-=\pm
\infty$, and we will draw the resulting diagrams with this in mind.

Having determined how regular flows navigate around singularities,
we can list the trajectories which can consistently be
extended to infinite RG time. 
These have been checked numerically, implying that they
are generic enough. The flows are depicted in Figures 1-3.  
We classify them according to where they start from and
where they terminate. The initial and final locations will either be
$g_+=0$ or $g_+=\infty$ as it takes an infinite time for
the flows to reach these locii.
For each flow there is a dual one,
and we only list one of the two flows in the dual pairs.
It is possible to draw the flows unambiguously in the $g_+, g_-$ plane
since $g_a$ flows in a simple manner:  in the UV $g_a=+\infty$ and
in the IR $g_a = -\infty$. 

\bigskip

\noindent {\bf Flows originating from the $g_+ = 0$ plane:}

\noindent
(a-1)  The trajectory returns to $g_+ = 0$  at some other point 
without crossing the singularities and
without $g_-$ going to $\infty$.  (Figure 1.) 

\noindent
(a-2) The flow is to  $g_- , g_a = \infty$ with $g_+<2$ 
without crossing the self-dual points. 
After identification of points at infinity as in eq.(\ref{patch1}),
the flow then continues at $g_-=-\infty$  
and again eventually returns  to the $g_+ = 0$ plane. 
(Figure 2.) 


\noindent
(a-3) The flow begins with $0<g_-<2$ and first flows towards the
self-dual point $g_\pm=2$, $g_a=0$. It goes through this point
according to eq.(\ref{through3}). After crossing the trajectory
is again determined by duality, and flows to $g_+ = \infty$.  (Figure 3.) 


\bigskip

\noindent {\bf Flows originating from the $g_+ = \infty$ plane, 
dual to those terminating at $g_+=0$:}

\noindent
(b-1) The flow starts at $g_+=\infty$ and loops back to $g_+=\infty$
without crossing singularities.  (Figures 1,3)  The flow of this kind
in Figure 3 is the dual of (a-2), and those in figure 1 the dual of (a-1).  

\noindent
(b-2) The flow  goes to $g_-,g_a=\infty$ without crossing
singularities.
After the identification (\ref{patch1}) the flow goes through the 
$g_\pm = \pm 2$ poles according to eq. (\ref{through1}) and then to
the $g_+ = 0$ plane.  (Figure 2.)  

\noindent
(b-3) The flow starts at infinity with $0<g_-<2$, goes up
to the self-dual point $g_\pm=2$, $g_a=0$, crosses it according 
 to (\ref{through3}) and terminates at $g_+=0$.  (Figure 2.) 

\noindent 
(b-4) The flow starts at infinity with $g_-<-2$, goes through
the lower self-dual point at $g_\pm=\pm2$, $g_a=0$,
according to (\ref{through1}), and then
continues up to $g_+=0$. It is a self-dual trajectory.  (Figure 2.) 

\bigskip

All of the above flows  asymptotically reach either
$g_+=0$ or $g_+=\infty$ at large scales.
They take an infinite time to arrive at these locii.
We shall refer to these phases as $dXY$ and $gX$ phases, respectively.
The flows for $g_+ < 0$ are the reflection of the flows for $g_+ >0$
about the $g_-$ axis due to the symmetry of the $\beta$ function. 


\medskip
{\underline {$\dXY$ phase}}:
$g_+ \to 0$, $g_-$ flows to a non-universal constant,
and $g_a$ flows to $-\infty$ logarithmically, 
$g_a\approx -a \log r$ with $a=4g_-^2/(2-g_-)^2$.
Since the non-zero couplings ($g_- , g_a$)  in the deep IR 
couple only the currents of the $gl(N|N)_1$ sub-current algebra 
this phase is in the same universality class as the disordered
$XY$ model, or Gade-Wegner class, as described in \cite{GLL}.  
Due to the structure of the $\beta$ functions (\ref{dxy}) since
$g_-$ does not flow to $\infty$, one cannot argue that this
phase flows to the fixed point $osp(2N|2N)_1/gl(N|N)_1$ as
the solution of the Virasoro master equation would suggest,  
and it remains unclear whether one can
identify a true IR fixed point to this flow.   
We point out 
that  it was shown in \cite{GLL} that
the $c=-2$  coset conformal field theory
$osp(2N|2N)_1/u(1)\otimes  u(1)$ describes a conformal sector but precise
arguments identifying this with the fixed point are missing.


\medskip
{\underline {$gX$ phase}}:
Here, 
$g_-$ flows  to a constant,
$g_a$ flows logarithmically to $-\infty$, $g_a\approx -a \log r$ with $a=4g_-^2/(2-g_-)^2$
whereas $g_+$ grows much faster as $g_+\approx \exp(2g_a^2/g_-^2)$.
Since all the $osp(2N|2N)$ currents are involved in this flow, this is probably
a massive phase.


\begin{figure}[htb]
\includegraphics[width=12cm]{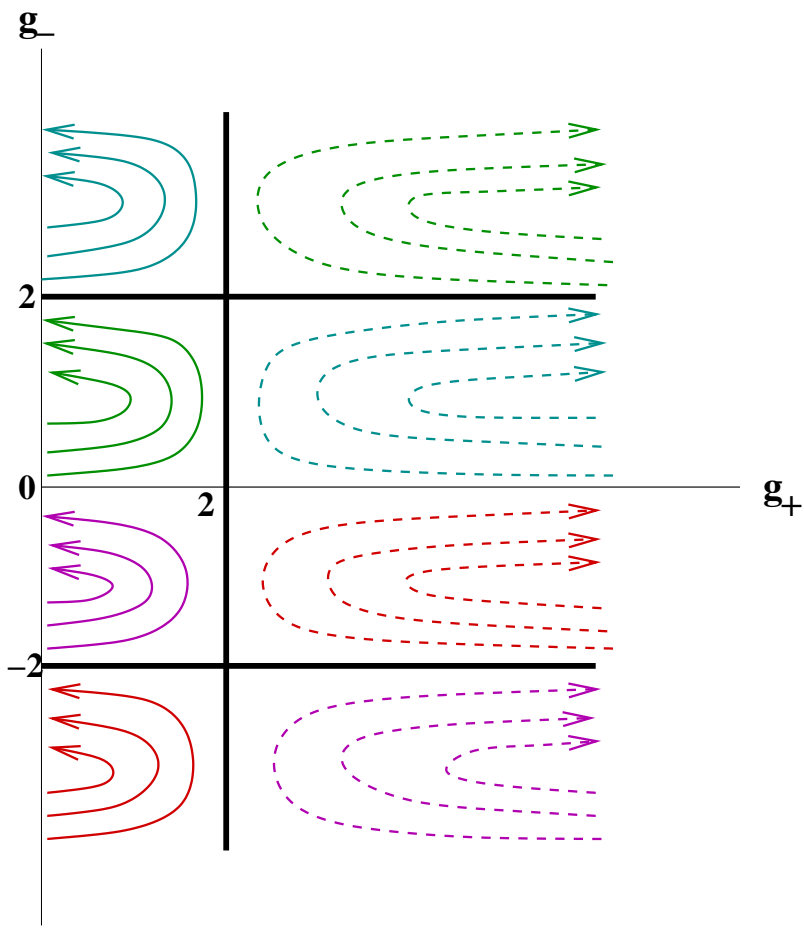}
\caption{RG trajectories that loop back to $g_+ = 0, \infty$ 
and flow to the $\dXY$ and $gX$  phases respectively. }  
\label{Figure1}
\end{figure}

\begin{figure}[htb]
\includegraphics[width=12cm]{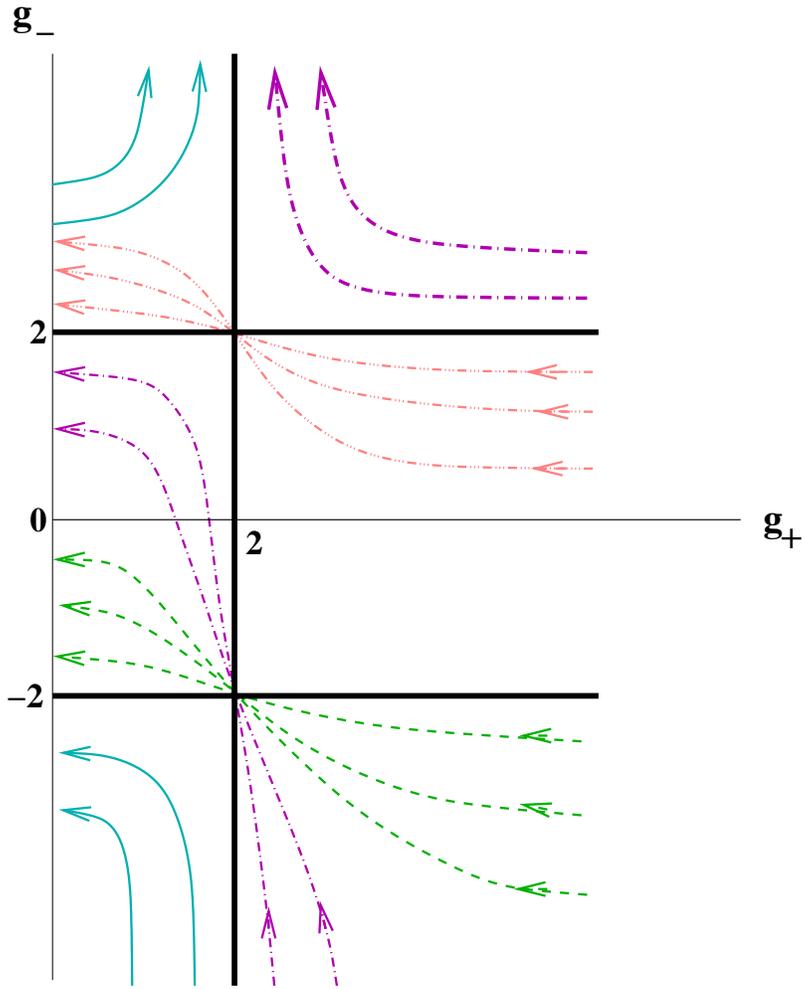}
\caption{RG trajectories that cross singularities and  flow to the $\dXY$ phase.} 
\label{Figure2}
\end{figure}

\begin{figure}[htb]
\includegraphics[width=12cm]{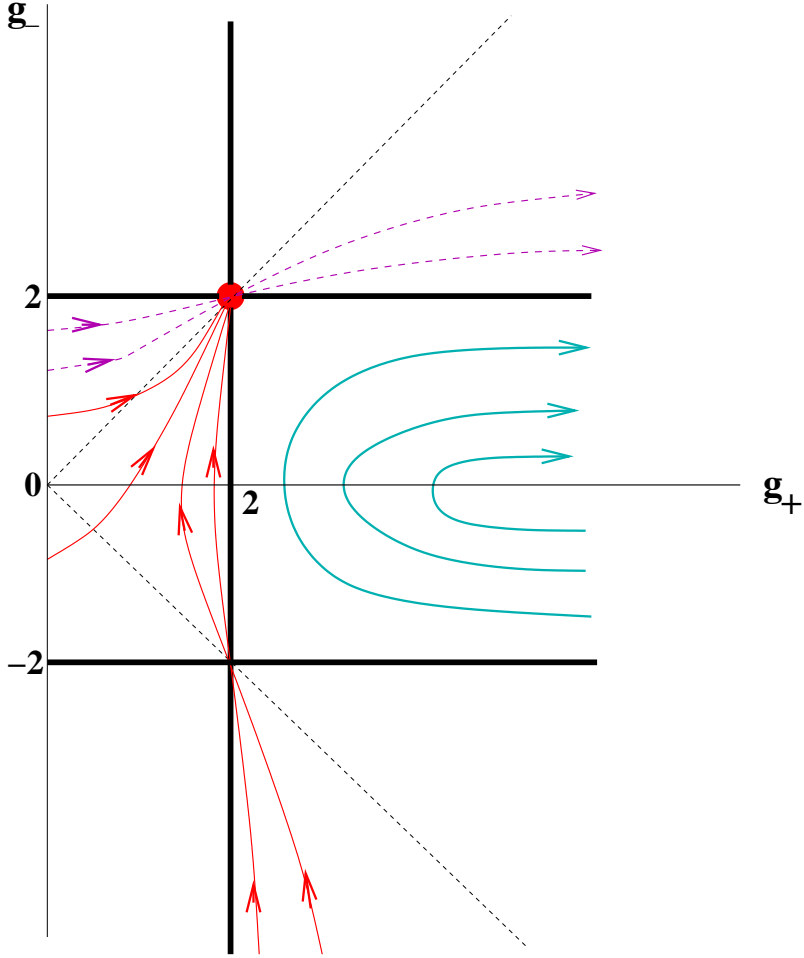} 
\caption{RG trajectories that either cross singularities and 
flow to $gX$ phase or are singular. 
The physical domain is the $90^\circ$ cone opening to the right. 
(Singular flows shown in red.) }
\label{Figure3}
\end{figure}

\section{The  spin quantum Hall network model}

\def\gm{g_m}

The hamiltonian of the network model for the spin quantum Hall
transition is 
\beq
\label{2.1s}
H = \left(\matrix{ 2\alvec\cdot\sigvec + m & -i \d_\zb + A_\zb 
         \cr -i \d_z + A_z & 2\alvec\cdot \sigvec - m \cr} \right) 
\eeq
where
$\alvec , m$ are random potentials
and  $A$ are random $su(2)$ gauge potentials\cite{sqhe,sqhe2,percolation}.
Our conventions for the Pauli matrices are 
\beq
\label{2.2}
\sigma^3 = \inv{2} \left( \matrix{1&0\cr0&-1\cr} \right) , 
~~~~
\sigma^+ = \inv{\stwo} \left( \matrix{0&1\cr0&0\cr} \right) ,~~~~~ 
\sigma^- = \inv{\stwo} \left( \matrix{0&0\cr1&0\cr} \right)  
\eeq
With the above convention, $Tr \sigma^a \sigma^b = \delta^{ab}/2$
and $f^{abc}f^{abd} = -2 \delta^{cd}$ where 
$[\sigma^a, \sigma^b]= f^{abc} \sigma^c$.  The random potentials,
$m,\ \alvec,\ A$ are taken to have the centered gaussian distributions
with variance $g_m,\ g_\al,\ g_s$.

\def\dx{ \frac{d^2 x}{2\pi} }


\def\betab{\bar{\beta}}

\subsection{Effective action and its  symmetries}

Introducing bosonic ghost partners $\beta_\pm , \bbar_\pm$ to
the fermions, the effective action upon disorder averaging is
\beq
\label{2.4s}
S_{\rm eff} = S_{\rm free } + \int \dx 
\( \gal \CO^\alpha + \gm \CO^m + g_s \CO^s \) 
\eeq
where 
\begin{eqnarray}
\nonumber
\CO^\al &=&  \( \psi_- \sigvec \psib_+ + \psib_- \sigvec \psi_+
+ \beta_- \sigvec \betab_+ + \betab_- \sigvec \beta_+ \)^2\\
\CO^m &=& \inv{4} \( \psi_- \psib_+ - \psib_- \psi_+ + \beta_- 
\betab_+ - \betab_-\beta_+ \)^2
\label{2.5s}
\\ 
\nonumber
\CO^s &=&  \(\psi_- \sigvec \psi_+ + \beta_- \sigvec \beta_+ \)
 \cdot\(\psib_- \sigvec \psib_+ + \betab_- \sigvec \betab_+ \)
\end{eqnarray}
In order to more clearly display the sub-algebraic structure,
let us define new couplings
\beq
\label{2.6s}
\gal \CO^\alpha + g_m \CO^m = g_c \CO^c + g_\o \CO^\o, ~~~~~
g_\alpha = g_c + \inv{2} g_\o , ~~~~g_m = \frac{3}{2} g_\o  - g_c  
\eeq
where $\CO^c = \CO^\alpha - \CO^m$,  
and $ \CO^\o = \inv{2} \CO^\alpha + \frac{3}{2} \CO^m $.
The  physical domain is the intersection of $g_c > -g_\o /2$
and $g_c< 3 g_\o /2 $ with $g_s > 0$.  
The free conformal field theory has a maximal
$osp(4|4)_{k=1}$ current algebra symmetry.
The Lie superalgebra $osp(4|4)$  has two commuting
subalgebras $osp(2|2)$ and $su(2)$.  
The above notation reflects the fact that $g_s$ ($g_c$) 
couples the spin (charge) degrees of freedom.   
Let us describe the $osp(4|4)$ currents in a notation
that clarifies these subalgebras.  
Define bosonic and fermionic $su(2)$ currents
\beq
\label{2.8s} 
L^a_\psi = \psi_- \sigma^a \psi_+ , ~~~~~~~~
L^a_\beta = \beta_- \sigma^a \beta_+ 
\eeq
These satisfy level $k=1$ and level $k=-1$ current algebra 
respectively
\barray
L^a_\psi (z) L^b_\psi (0) \sim \frac{1}{z^2} 
\frac{\delta^{ab}}{2} + \inv{z} f^{abc} L^c_\psi (0) \\
\label{2.9s}
L^a_\beta (z) L^b_\beta (0) \sim - \frac{1}{z^2}
\frac{\delta^{ab}}{2} + \inv{z} f^{abc} L^c_\beta (0) 
\earray
The operator $\CO^s$ is the $su(2)$ invariant built from the
sum:
\beq
\label{2.10s}
\CO^s = \sum_a  \CL^a \bar{\CL}^a , ~~~~~~~~~~
\CL = L_\psi + L_\beta
\eeq
The current $\CL^a$ satisfies a level $k=0$ current algebra.  

\def\Jb{\bar{J}} 
\def\Sh{\hat{S}}
\def\Hb{\bar{H}}
\def\Sb{\bar{S}}

The currents in $\CO^c$ generate an $osp(2|2)$ subalgebra.
Define
\begin{eqnarray}
\nonumber
J &=& \sum_i  \psi_+^i \psi_-^i  ,  \quad
J_\pm  = \sum_{i,j} \ep_{ij} \psi^i_\pm \psi^j_\pm
\\ \label{2.11s}
H &=& \sum_i \beta_+^i \beta_-^i, \quad
S_\pm =  \sum_i \psi_\pm^i \beta_\mp^i,\quad
\hat{S}_\pm  = \sum_{i,j}  \ep_{ij} \psi^i_\pm \beta^j_\pm 
\nonumber
\end{eqnarray}
where $\ep_{ij} = -\ep_{ji}$, $\ep_{12} = 1$, 
and similarly for the 
right-movers
$\bar{J} = \psib_+ \psib_-,  \bar{H} = \betab_+ \betab_-, .....$.
These currents generate an $osp(2|2)_k$ 
current algebra at level $k=-2$\cite{BLsqhe}.
The currents $J$ and $J_\pm$ generate a charged $su(2)$ subalgebra under
which the fermions $\psi^i_\pm$ transform as doublets.

The operator $\CO^c$ is the $osp(2|2)$ invariant bilinear:
\beq
\label{2.13s}
\CO^c 
=  - J\Jb + H\Hb + \inv{2} (J_- \Jb_+ + J_+ \Jb_- )
+ S_- \Sb_+ - S_+ \Sb_-  + \Sh_+ \bar{\Sh}_- 
-\Sh_- \bar{\Sh}_+ 
\eeq
To show this we have used the identity (\ref{identi}) of the Appendix.

\def\Vb{\bar{V}}
\def\Lb{\bar{L}}
\def\rhat{\hat{\rho}}
\def\Ub{\bar{U}}
\def\Vb{\bar{V}}
\def\Bb{\bar{B}}

The currents in $\CO^\o$ are orthogonal to the above ones.  
To better reveal the $su(2)$ properties of these  currents,
let us  introduce 
intertwiners between the symmetric tensor product
of two spin $1/2$ representations and the spin $1$ representation.
These are:
\beq
\label{2.16}
\rho^a = \sigma^a \ep, ~~~~~
\hat{\rho}^a = \ep \sigma^a,  ~~~~~
\ep = \left(\matrix{0&1\cr-1&0\cr}\right) 
\eeq
We therefore define
\barray
\label{2.18}
B_+^a &=& \inv{\stwo} \beta_+ \rhat^a \beta_+ , ~~~~~
B_-^a = \inv{\stwo} \beta_- \rho^a \beta_- \\
U_+^a &=& \beta_+ \rhat^a \psi_+, ~~~~~
U_-^a = \beta_- \rho^a \psi_-
\\
V_-^a &=& \psi_- \sigma^a \beta_+ ,~~~~~
V_+^a = \beta_- \sigma^a \psi_+ 
\earray
They transform in the spin $1$ representation of $su(2)$. 
The operator $\CO^\o$ can then be written as 
\beq
\label{2.20}
\CO^\o = -2 \( B_- \Bb_+ + B_+ \Bb_- + U_- \Ub_+ - U_+ \Ub_- 
+ V_- \Vb_+ - V_+ \Vb_-  +L_\beta \Lb_\beta - L_\psi \Lb_\psi \) 
\eeq
where e.g. $B_- \Bb_+ = \sum_a B_-^a \Bb_+^a$.  

We can now describe the $osp(2|2)\otimes su(2)$ transformation
properties of the above currents.  The $osp(2|2)$ currents
(\ref{2.11s}) 
commute with the $su(2)$ currents $\CL^a$.  
The currents in $\CO^\o$ transform in the spin $1$ representation
of $su(2)$. 
The fields $L_{\psi, \beta}$ are not primary however:
\barray
\label{2.22s}
\CL^a (z) L^b_\psi (0) &\sim& \inv{z^2} \frac{\delta^{ab}}{2} 
+ \inv{z} f^{abc} L^c_\psi 
\\ \nonumber
\CL^a (z) L^b_\beta (0) &\sim& -\inv{z^2} \frac{\delta^{ab}}{2}
+ \inv{z} f^{abc} L^c_\beta
\earray

Under the $osp(2|2)$ symmetry the currents in $\CO^\o$ transform in
the $8$ dimensional indecomposable representation. 
This can be expressed as the OPE
\beq
\label{2.23s} 
S^\alpha (z) J^{i,a}_\o (0) \sim \inv{z} 
(t^\alpha)^i_j J^{j,a}_\o 
\eeq
where $S^\alpha$ are the $osp(2|2)$ currents, 
$J^{i,a}_\o$, $i=1,..,8$, $a=1..3$ are the $24$ currents in
$\CO^\o$ with $a$ the $su(2)$ index, and $t^\alpha$ is 
an $8$ dimensional matrix representation of $osp(2|2)$.  

\def\Ct{\tilde{C}}

\section{The $\beta$eta function for the spin network model}

The interactions that perturb away from the conformal
field theory are of the form (\ref{action}) 
where $J^a$ are the $osp(4|4)$ currents at level $k=1$ described
above, and $d^A_{ab}$ are quadratic forms that define
the operators $\CO^A = \{ \CO^s , \CO^c, \CO^\o \}$.
As before, the $\beta$eta function proposed in \cite{GLM} is expressed
in terms of the data $C^{AB}_C , D^{AB}_C, \Ct^{AB}_C$ 
which can be computed from the OPE's (\ref{2.10s}).   
One finds the non-zero values
\barray
\nonumber
C^{ss}_s &=&  C^{s\o}_\o = C^{\o s}_\o = -2 \\ \nonumber
C^{cc}_c &=& C^{\o\o}_\o = 4, \quad  C^{\o\o}_c = -3\\\label{3.4} 
C^{c\o}_s &=& C^{\o c}_s = C^{\o\o}_s = -8 
\earray
The coefficients $C^{ss}_s$ and $C^{cc}_c$ are easily
understood as minus the quadratic Casimir in the adjoint
representation of $su(2)$ and $osp(2|2)$ respectively.

The other RG data involves OPE's of the purely left-moving
operators (\ref{3.5b}).
One can easily disentangle the $D$ from the $\Ct$ term by
noting that the $D$ term only arises from central terms
proportional to $k$ in the OPE.  
One finds 
\beq
\label{3.7} 
D^{cc}_c = -2, ~~~~~D^{\o s}_s = D^{s\o}_s =  D^{\o\o}_\o = 1  
\eeq
Some of the $\Ct$ coefficients follow from simple group theory. 
 Consider first OPE's with $T^s$. 
Since $T^s$ has the form of the quadratic Casimir for
$su(2)$, we have $T^s \CO^s \sim 2 \CO^s $, 
$T^s \CO^\o \sim 2 \CO^\o$ since $2$ is the quadratic
Casimir for the spin $1$ representation of $su(2)$. 
Thus $\Ct^{ss}_s = \Ct^{s\o}_\o = 2$.   
Similarly $T^c \CO^c \sim -4 \CO^c$ where $-4$ 
is the Casimir for the adjoint representation of $osp(2|2)$. 
The only non-zero OPE's of $T^c$ with $\CO^\o$ currents 
are 
\beq
\label{3.8}
T^c (z) L_\psi \sim - T^c (z) L_\beta \sim \frac{4}{z^2} \CL
\eeq
which implies $\Ct^{c\o}_s = 8$.  The remaining $\Ct$ are
computed  tediously but straightforwardly using the  free-field   OPE's
(\ref{2.6}).     For convenience we collect all non-zero
$\Ct$
\barray
\nonumber
\Ct^{ss}_s &=& \Ct^{s\o}_\o = 2 \\\nonumber
\Ct^{\o\o}_s &=& -\Ct^{\o s}_s = -\Ct^{cc}_c =-\Ct^{\o\o}_\o  = 4\\
\label{3.9} \Ct^{c\o}_s &=&  8, \quad  \Ct^{\o c}_c = -6 
\earray

%

Using the general expression in 
\cite{GLM}, the result for the beta functions is 
\barray
\nonumber
\beta_{g_\o} &=& \frac{-8 g_\o}{(g_\o+2)^3 (g_\o-2)} 
\( g_\o^3 + g_\o^2 g_s - 4 g_\o + 4 g_s \) \\ 
\label{betas}
\beta_{g_c} &=& \frac{2}{(g_\o^2 - 4)^2 (g_c -1 )^2 } 
\( 12 g_\o^2 g_c^4 - g_\o^4 g_c^2 - 16 g_\o^2 g_c^2 - 16 g_c^2 + 12 g_\o^2\) \\
\nonumber
\beta_{g_s} &=& 
\frac{-4}{(g_c-1)(g_\o-2)^2(g_\o+2)^4} 
\BL g_s^2(g_c-1)(3g_\o^4 + 4g_\o^3 - 8g_\o^2 + 16 g_\o - 16) \\
\nonumber
&&~~~~~~~~+ 4g_\o^2 g_s(g_\o-2)(g_\o+2)^2 (g_c-1) 
-2 g_\o(g_\o-2)(g_\o^2+4)(g_\o+2)^2(g_\o+2g_c) \BR
\earray
The \betaf function remarkably also has the duality (\ref{2.12}) 
with 
\beq
\label{F1}
g_c^* = \inv{g_c}, ~~~~~g_\o^* = \frac{4}{g_\o} ~~~~~
g_s^* = - \frac{4 g_s}{g_\o^2}
\eeq
The self-dual points are $(g_s, g_c, g_\o) = (0, \pm 1 , \pm 2)$.  
As for the $u(1)$ network model, we have not found any RG
invariants, unlike the $su(2)$ case.

\subsection{Virasoro master equation and critical exponents} 

As for the $u(1)$ network model,  let us describe  the solutions of the 
Virasoro master equation 
(\ref{master}).
We obtain four pairs of solutions.
First one is the trivial one:
$$
T_0=0 \quad {\rm and} \quad T_{osp(4|4)_1} =\inv{4}T^c-\inv{2}T^\o
$$
Two other pairs of solutions are given by:
\debut
T_{osp(2|2)_{-2}} =-\inv{8} T^c \quad &{\rm and}& \quad
T_{osp(4|4)_1 / osp(2|2)_{-2} } = \frac{3}{8} T^c -  \inv{2}T^\o \non\\
T_{su(2)_0} = \inv{2} T^s \quad &{\rm and}& \quad
T_{osp(4|4)_1/su(2)_0} =-\inv{2}T^s +\inv{4}T^c -\inv{2}T^\o \non
\fin
The last pair of solutions,
$$
T=\inv{2}T^s -\inv{8}T^c \quad {\rm and} \quad
T' =-\inv{2}T^s +\frac{3}{8}T^c-\inv{2}T^\o
$$
corresponds to $osp(2|2)_{-2}\otimes su(2)_{0}$
and to the coset $osp(4|4)_{1}/osp(2|2)_{-2}\otimes su(2)_{0}$.

Let us recall the spin-charge
separation of the free conformal field theory of the bosons and
fermions $\psi , \beta$.  This free theory has
$osp(4|4)_1$ current algebra symmetry with stress tensor
$T_{osp(4|4)_1}$.   It was shown in Ref. \cite{BLsqhe} that
the latter stress tensor decomposes as follows:
\beq
\label{fp1}
T_{osp(4|4)_1} = T_{osp(2|2)_{-2}} + T_{su(2)_0}
\eeq
where $T_{osp(2|2)_{-2}}$ and $T_{su(2)_0}$ are the affine-Sugawara
stress tensors for the $osp(2|2)_{-2} $ and $su(2)_0$ current algebras.
This lead to the identity $T'=0$:
\beq
\label{idents}
3 T^c - 4T^8 - 4 T^s  = 0 
\eeq

It was argued  in Ref. \cite{BLsqhe} that though the stress tensor decomposes
as (\ref{fp1}), the Hilbert space does not exactly factorize due
to logarithmic pairs such as $\CO^\o , \CO^s$.
This means that the coset $osp(4|4)_1 / su(2)_0$ is not
identical to the current algebra $osp(2|2)_{-2}$ even though it
has a global $osp(2|2)$ symmetry and the scaling dimensions are
correctly given by $osp(2|2)_{-2}$.
However recent work shows that 
certain 4-point correlation functions can be factorized\cite{Bhaseen}.
A comparison with other approaches was studied in \cite{Tsvelik2}. 

The interesting possible fixed points are thus the cosets
${osp(4|4)_1}/{su(2)_0}$ and ${osp(4|4)_1}/{osp(2|2)_{-2}}$.   
For each fixed point, let us  compute some of the
critical exponents.  The density of states is proportional
to the one-point function of $\Phi_\CE$,
\beq
\label{dens}
\bar{\rho (E) } \propto \langle \Phi_\CE \rangle
\eeq
where $\Phi_\CE$ is defined in eq. (\ref{phiE}).
Let $\Gamma_E$ denote the scaling dimension of $\Phi_{\CE}$
in the IR.   Then
\beq
\label{den2}
\bar{\rho (E)} \propto E^{\Gamma_E/(2-\Gamma_E)}, ~~~~~~~
{\rm as} ~ E \to 0
\eeq
Other exponents are generally related to other perturbations
$\Phi_\delta$:
\beq
\label{perts}
\delta S = \delta \int d^2 x  ~ \Phi_\delta
\eeq
where $\delta = 0$ is the critical point.  Letting
$\Gamma_\delta$ denote the scaling dimension of $\Phi_\delta$
in the IR,  the mass dimension of $\delta$ is then $2-\Gamma_\delta$.
As $\delta \to 0$, there is thus a diverging length $\xi$,
\beq
\label{xi}
\xi \propto \delta^{-\nu}  , ~~~~~\nu = \inv{2-\Gamma_\delta}
\eeq
For a coset fixed point $\CG_k/ \CH_{k'}$,
 the dimensions of the fields in the IR
are given by
\beq
\label{dims}
\Gamma = 2\Delta
= 2 \(  \Delta^{\CG_k}  - \Delta^{\CH_{k'}} \) 
\eeq

The spin-charge separation implies
that the conformal scaling dimensions $\Delta$ satisfy
the sum rule:
\beq
\label{sumrule}
\Delta^{osp(4|4)_1} = \Delta^{osp(2|2)_{-2}} + \Delta^{su(2)_0}
\eeq
The $osp(4|4)$ scaling dimensions are simply given by the free theory
and are thus integer or half-integer, where
$\psi , \beta$ have $\Delta^{osp(4|4)_1} = 1/2$.  The primary fields of the
$osp(2|2)_{-2}$ can be labeled $(j,b)$ where $j$ is the spin of the
$su(2)$ generated by $J, J^\pm$ and $b$ is the $u(1)$ charge $H/2$.
These fields form $8j$-dimensional multiplets and have left/right
scaling dimension\cite{Serban} 
\beq
\label{ospdel}
\Delta^{osp(2|2)_{-2}}_{(j,b)}  = \frac{j^2 - b^2}{2}
\eeq
The primary fields of $su(2)_0$ on the other hand are labeled by
a spin $j$ and have scaling dimension\cite{KZ}
\beq
\label{su2dim}
\Delta^{su(2)_0}_j  = \frac{j(j+2)}{2}
\eeq

\underline{$osp(4|4)_1/osp(2|2)_{-2}$}  
The $osp(2|2)_{-2}$ dimension of $\Phi_\CE$ is $1/4$ since
it transforms in the $(1/2, 0)$ representation with $\Delta^{osp(2|2)}
 = 1/8$.
Thus $\Gamma_E = 1- 1/4= 3/4$ and one finds
\beq
\label{densc}
\bar{\rho (E)} \propto E^{3/5}
\eeq
In this phase it is the charge degrees of freedom that are massive,
i.e. localized.
 
\bigskip
\underline{$osp(4|4)_1/su(2)_0$}.
The field $\Phi_E$ has $j=1/2$ quantum numbers under the $su(2)_0$,
and thus has $\Delta^{su(2)} = 3/8$.   This leads to $\Gamma_E = 1-3/4= 1/4$
and
\beq
\label{denss}
\bar{\rho (E)} \propto E^{1/7}
\eeq

Let us compare the latter coset   with percolation.
The map to percolation involves averaging over $su(2)$ matrices
with the flat Haar measure, which is expected to correspond to $g_s = \infty$.  
The $g_s = \infty$ theory was described in \cite{Tsvelik}. 
This averaging projects out the $su(2)$ degrees of freedom and in our
approach this is analogous to arguing that the RG flow gaps out the
$su(2)_0$ sub-current algebra.  
We do not have an argument leading to the identification of
 $\Phi_\delta$ for the localization length exponent.  However
there does exist a field in our theory with the correct dimension
to lead to the percolation exponent.
Consider  a field $\Phi_\delta$ in the free theory of the
form $\Phi_\delta = \phi_\delta \bar{\phi_\delta}$ with
$\phi_\delta \sim (\psi\psi\psi\beta\beta)$.  In the
free theory this field has $\Delta = 5/2$.  Since $\psi, \beta$
have $j=1/2$ under the $su(2)_0$, we can take $\phi_\delta$
to be in the $j=3/2$ representation of $su(2)_0$ with
$\Delta^{su(2)_0} = 15/8$.  Under the $osp(2|2)_{-2}$
this field transforms in the $(j,b) = (3/2, 1)$ representation
with $\Delta^{osp(2|2)} = 5/8$.   Note that the sum rule
is satisfied:  $15/8 + 5/8 = 5/2$.   
Since $\Gamma_\delta = 5/4$ this  leads to $\nu = 4/3$.

\subsection{RG flows}

In this section we describe the RG flows based on  
the \betaf function (\ref{betas}).   As in the $u(1)$ network model,
there exist both well-behaved and singular flows. 
 
\bigskip

\noindent
\underline{General properties.}

When $g_\o=0$ the spin and charge degrees of freedom decouple
and the \betaf function reduces to
\beq
\label{A1}
\beta_{g_s} = g_s^2 , ~~~~~~ \beta_{g_c} = - \frac{2 g_c^2}{(g_c-1)^2}
\eeq
The fact that the one-loop result for $\beta_{g_s}$ is exact
is due to the level $k$ being zero for the $su(2)$ currents
$\CL^a$.   
If $g_s >0$, $g_s$ flows to infinity after a finite scale
transformation, whereas if $g_s <0$ it flows to zero, i.e. 
$g_s>0$ is marginally relevant and $g_s<0 $ is marginally irrelevant. 
On the other hand $g_c <0$ is marginally relevant with $g_c$ flowing
to $-\infty$ and $g_c >0$ is marginally irrelevant with $g_c$ flowing
to zero.  

When $g_\o \neq 0$, we can consider the flows as originating
from the $g_s -g_c$ plane.  On this plane the flows originate
in the UV from $g_s = 0 , -\infty$ or $g_c = 0,+\infty$.
Near $g_\o = 0$, $\beta_{g_\o} = 2 g_s g_\o$.  
Thus if $g_s >0$ $(g_s <0)$ one flows away (toward) the 
$g_s - g_c$ plane.  If $g_s$ is not large enough, the flows
can turn around and form a loop returning to the $g_\o = 0$ plane.
Otherwise it continues away from the plane until it reaches
a singularity.   
By duality there are similar properties at $g_\o = \infty$.

At generic points, the slope of the trajectories 
$\d g_\o / \d g_s = \beta_{g_\o} /\beta_{g_s}$ is zero at $g_c=\pm 2$,
whereas  the slope $\d g_\o / \d g_c = \beta_{g_\o} /\beta_{g_c}$ is zero at
$g_\o =2$ but equals $\infty$ at $g_\o = -2$.  
Thus the flows are generically  tangent to the $g_\o = 2$ plane,
since $\beta_{g_s},\ \beta_{g_c}\gg \beta_{g_\o}$ near $g_\o = 2$. 
Furthermore, $\beta_{g_c}\simeq 6 (g_c^2-1)^2/(g_\o -2)^2$, so that
$g_c$ increases in the vicinity of the plane $g_\o=2$.
On the other hand, since $\beta_{g_c}\ll \beta_{g_\o}\ll \beta_{g_s}$
near $g_\o = -2$ on the other hand, the flows are generically attracted
or repealed by the $g_\o=-2$ plane. More precisely, 
because of the odd order pole,
$\beta_{g_\o} \simeq -32 g_s/(g_\o+2)^3$,
the flows are attracted to $g_\o = -2$ for $g_s >0$,
whereas for $g_s <0$ flows are away from $g_\o = -2$.  
One also has $\beta_{g_s} \simeq 16g_s^2/(g_\o +2)^4$ 
near $g_\o = -2$, the coupling $g_s$ very rapidly increases.

The behavior near the poles $g_c = 1$ is somewhat simpler.
There, the slopes are such that $\d g_\o / \d g_c$  and
$\d g_s / \d g_c$ both vanish since
 $\beta_{g_\o}\ll \beta_{g_s}\ll\beta_{g_c}$ at generic points.
Near $g_c = 1$ and away from $g_\o = \pm 2$ the beta function
$\beta_{g_c} \simeq -2/(g_c -1)^2$ has double pole. 
Thus $g_c$ always decreases and flows can cross the $g_c =1$ plane.  
Note that $g_c=-1$ is not a singular point.
Near $g_c=-1$ one finds $\beta_{g_c} = -1/2$, 
so again the flows easily cross the $g_c = -1$ plane.  

The non generic singular points are those for which
the ratio of the beta functions is finite and non-vanishing.
These are at $(g_0=2,g_c=\pm1)$.
At $g_c = -1, g_\o =2$ and $g_s = 0$
the ratio $\d g_\o / \d g_c$ and $\d g_\o / \d g_s$ are finite.
Thus one can flow through this self-dual point and
extend the flow by duality. 
We have numerically verified such flows.  
As analyzed below, 
the flows toward $(g_c , g_\o) = (1,2)$ on the other hand are singular.

As in previous examples, blow-ups in finite RG time $\tau$
are resolved by certain topological identifications.  
When $g_\o\neq 0$, there are flows that blow up in finite
RG time in the direction $g_c \to \infty$.
As can be seen numerically, in this domain $g_c$ increases
rapidly while $g_\o$ flows  close to the value $g_\o=2$.
This can be seen by expanding near $g_\o=2+\ep_\o$ with $\ep_\o\to 0$.
The $\beta$-functions asymptotically reduce to 
\beq
\label{as3}
\beta_s\simeq -\frac{g^2_s}{\ep_0},~~~~~
\beta_c\simeq \frac{6g_c^2}{\ep_\o^2}, 
~~~~
\beta_\o\simeq -\frac{2g_s}{\ep_\o}
\eeq
The solutions are 
\beq
\label{as4}
g_s \simeq \sqrt{ \ep_\o + {\rm const.}}
, ~~~~~
\ep_\o\simeq 2c\sqrt{(\tau_*-\tau)}, ~~~~~
g_c^{-1}\simeq \frac{3}{2c^2}  \log((\tau_*-\tau)b)
\eeq
showing that $(g_s , g_c , g_\o) \to ({\rm const.}, \infty, 2)$. 
Note also that $g_c$ blows up before $\ep_\o$ vanishes.
To continue the flows note that the asymptotic behavior
(\ref{as3}) satisfies
$\beta( g_s , g_c , g_\o)  =  \beta(g_s , -g_c , g_\o)$ 
as $|g_c| \to \infty$.   Viewing this as
a consistent patching condition, we can thus identify $g_c = +\infty$ with
$g_c = -\infty$ when $g_\o \neq 0$, such that flows toward 
$g_c = \infty$ continue at $g_c = -\infty$.

\bigskip
\noindent
\underline{Singular flows.} 
There are two classes of singular flows:

(i)  
Consider the flows toward the self-dual point
$g_c=1, g_\o=2$.  
Letting $g_c = 1 + \ep_c$, $g_\o = 2 + \ep_\o$,
asymptotically the \betaf function behaves as
\barray
\nonumber
\beta_{g_s} &\simeq& -\frac{g_s^2}{\ep_\o^2} + \frac{32}{\ep_c \ep_\o}
 + \frac{16- 4g_s-g_s^2}{\ep_\o} + \frac{24}{\ep_c}
+ \cdots
\\ \label{as8}
\beta_{g_c} &\simeq& \frac{24}{\ep_\o^2} - \frac{2}{\ep_c^2}
+ \frac{12}{\ep_\o} - \frac{4}{\ep_c} + \cdots
\\
\beta_{g_\o} &\simeq& -\frac{2g_s}{\ep_\o} + \cdots
\earray
The trajectories fine tune themselves
coherently in order to converge
toward the singularity and to cancel the double poles of the
$\beta$-functions.
The RG trajectories may be found as series expansion in terms
of $\ep_\o$, with $\ep_\o\to 0$.
We obtain
$$\ep_c=\ep_\o(1+ a_c\ep_\o+\cdots)/2\sqrt{3}$$
with $a_c$
solution of $3(4\sqrt{3} a_c+\sqrt{3}-1)/2=-3^{1/4}$ and
$$g_s  =8\cdot 3^{1/4} (1 + \inv{6\cdot 3^{1/4}} \ep_\o\log \ep_\o+\cdots)$$
The RG equation for $\ep_\o$, $\dot \ep_\o=\beta_\o$, then gives
$$\ep_\o\simeq - 3^{1/8}\sqrt{32(\tau_*-\tau)}+\cdots$$
showing that the
trajectories are effectively attracted by the pole.
Since $g_s$ flows to $8\cdot 3^{1/4}$ rather than zero, this
flow cannot be continued through the singularity by duality.  
These  singular flows are  in the physical
domain.

(ii)   Consider the flows toward $g_\o = -2$.  Letting
$g_\o = -2 + \ep_\o$, the \betaf function reduces to 
\beq
\label{as5}
\beta_{g_s} \simeq \frac{16 g_s^2}{\ep_\o^4} , ~~~~~
\beta_{g_c} \simeq \frac{6(g_c+1)^2}{\ep_\o^2} , ~~~~~
\beta_{\ep_\o} \simeq -\frac{32g_s}{\ep_\o^3} 
\eeq
Integrating these equations one finds the asymptotic behavior
\beq
\label{as6}
g_s \simeq c \ep_\o^{-1/2} , ~~~~~
\inv{g_c +1} \simeq \frac{3}{40c} \ep_\o^{5/2} + c' 
,~~~~~
\ep_\o \simeq (144c(\tau_* - \tau))^{2/9} 
\eeq
for some constants $c, c', \tau_*$.  Thus in a finite RG time 
$(g_s , g_c , g_\o) \to (\infty, {\rm const.}, -2)$.  
Again, there is neither the duality nor a consistent patching
condition allows the flows to extend beyond this singularity.

\bigskip
\noindent
\underline{Regular flows}
All other flows can be extended to arbitrarily large or small length
scales.  These well-behaved flows either don't encounter singularities
or can be extended consistently  by passing between patches  
patches glued at $g_c = \pm \infty$, and/or  flowing  through the 
self-dual points $(g_s, g_c , g_\o ) = (0, -1, 2)$.   
The flows that do not loop back to $g_\o =0$ or $\infty$ are shown in
Figures 4,5.   

Below we list the asymptotic directions,  or phases, that can be
continued to arbitrarily large length scales.

{\underline {phase $O$}}  ~~~ In this phase all disorder is
marginally irrelevant.  The flows are attracted to $g_\o = 0$ with
$g_s <0, g_c >0$ and then  $g_s , g_c$ continue to flow to zero 
according to the \betaf function (\ref{A1}).

\medskip
{\underline {phase $C$}} ~~~ This phase is similar to phase $O$ except  
that $g_c <0$ and thus marginally relevant.   Thus 
$g_s , g_\o \to 0$ and $g_c \to -\infty$. 

\medskip

{\underline {phases $S^\pm$}}  ~~~
A large set of flows, some with rather different UV properties,
asymptotically flow in the direction $S^+$.  
Some  flows originate in the UV with $g_c >0$.
Flows leaving the $g_c -g_s$ plane with $g_c >0$ flow to
$g_c = \infty$, continue at $g_c = - \infty$, through the 
self-dual point $(g_s , g_c, g_\o) = (0, -1, 2)$ and then in 
the $S^+$ direction.  
These flows originate from 
$(g_s , g_c , g_\o )_{UV} = (0^+ , -\infty , 0^+)$, and
then by duality should flow to $g^*$, i.e. 
in the IR we expect $(g_s , g_c , g_\o)_{IR} = (-\infty , 0 , +\infty)$. 
As shown below, the  asymptotic behavior  of $S^+$ supports this.   
Flows originating from $(g_s, g_c , g_\o) = (+\infty, 0, +\infty)$
are also attracted to $S^+$ but do not pass through self-dual points.

Assuming $g_\o g_c \gg 1$, the \betaf function asymptotically reduces to
$$\beta_\o\simeq- 8g_s/g_\o, ~~~~~\beta_c\simeq 2g_c^2, ~~~~~
\beta_s\simeq -12 g^2_s/g_\o^2 - 8g_\o$$
The equation for $g_c$
decouples and $g_c\simeq 1/\tau$ with $\tau$ the RG time.
Integrating the RG equations in the  $g_\o,~g_s$ plane gives
the equations for the asymptotic trajectories,
$$g_s^2\simeq 2 g_\o^3\log g_\o, 
~~~~ g_\o\propto \tau^2\log \tau, ~~~~~g_c \propto 1/\tau
$$
for large $\tau$. As a consistency check one verifies that
$g_\o^2\ll g_s^2$ and $g_\o g_c\gg 1$.  
The phase $S^-$ has the same properties as $S^+$ with 
$g_\o \to -\infty$.  

\medskip

{\underline {phase $SC^\pm $}}  ~~~  
These flows originate from the $g_c -g_s$ plane
with $g_c <0$.  In the deep UV one has 
$(g_s , g_c, g_\o)_{UV} = (0^+, 0^-, 0^+)$.  Since the 
flow is through a self-dual point, one expects from duality that
$(g_s , g_c , g_\o)_{IR} = (+\infty, -\infty, +\infty)$.  
Assuming  $|g_\o|\ll |g_s|$ and
$|g_\o^3|\ll |g_cg_s^2|$,  
the $\beta$-functions reduce to
$$\beta_\o\simeq -\frac{8g_s}{g_\o},~~~~~ \beta_c\simeq -2, ~~~~~
\beta_s\simeq -\frac{12 g^2_s}{g_\o^2}$$
The trajectories read 
$$ g_s^2 \simeq c g_\o^3 , ~~~~~ g_\o \simeq 16 c \tau^2 
, ~~~~ g_c \simeq -2\tau $$
for large $\tau$ where $c$ is some constant.
The phase $SC^-$ is very similar to $SC^+$ but with $g_\o <0$.

\begin{figure}[htb]
\hspace{18mm}
\includegraphics[width=15cm]{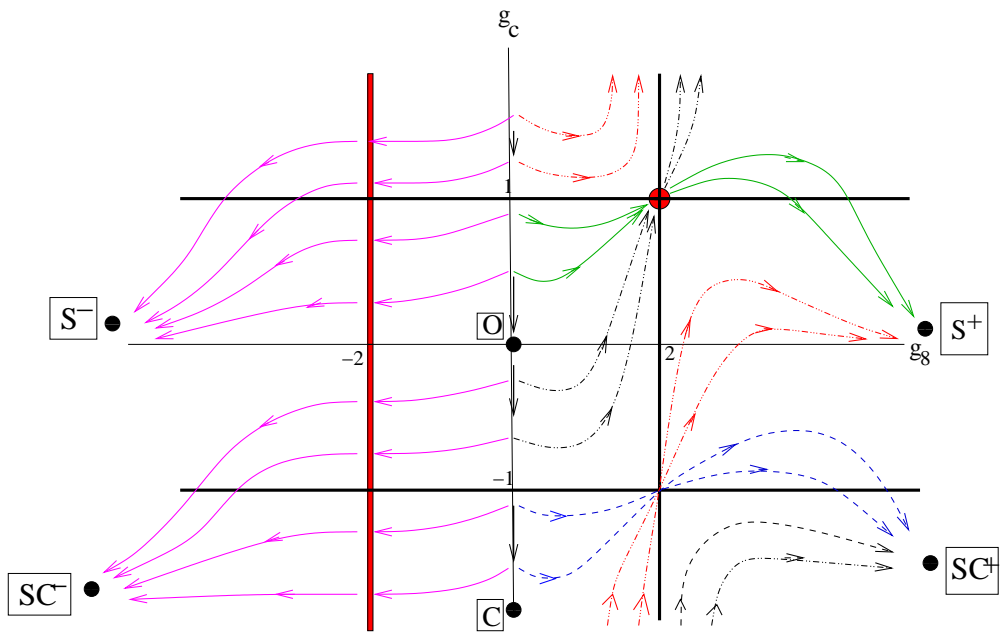}
\caption{Flows away from $g_\o = 0$ in the spin network model.} 
\label{Figure4}
\end{figure}

\begin{figure}[htb]
\hspace{18mm}
\includegraphics[width=15cm]{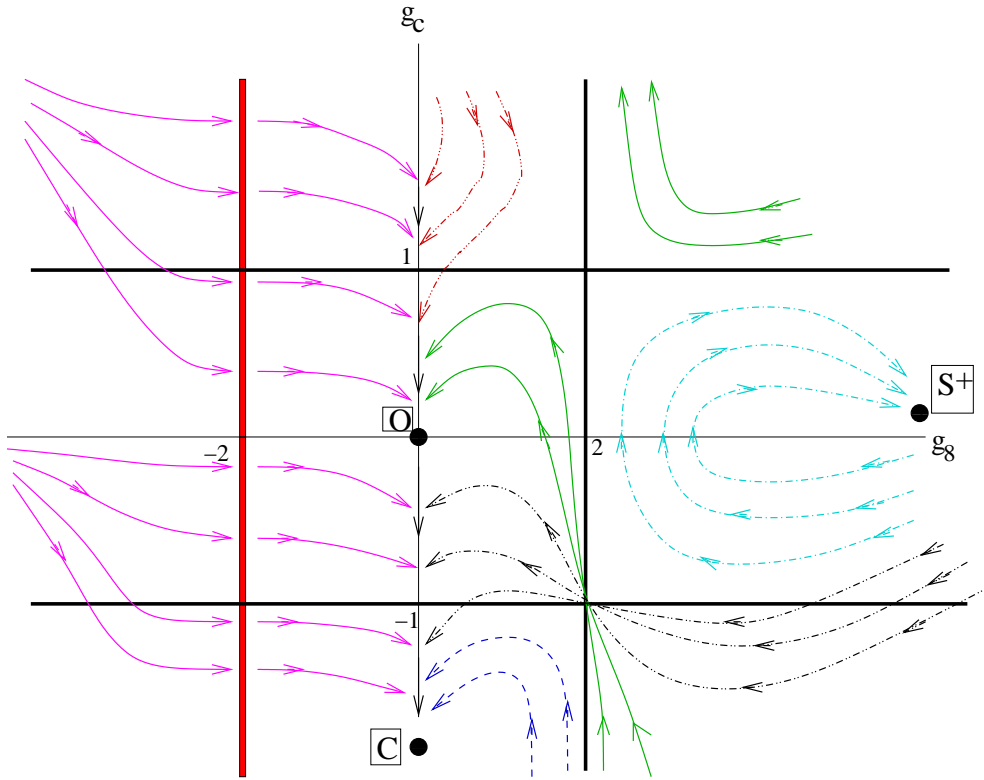}
\caption{Flows toward $g_\o =0$ in the spin network model.}
\label{Figure5}
\end{figure}

\subsection{Identifying the fixed points}

It is possible to interpret 
the asymptotic directions listed in the previous subsection
as some of the above coset solutions of
the master equation. 
The main hypothesis in making this identification is the following. 
Suppose a coupling $g$ flows to infinity much faster than the other couplings
and $g$ couples a sub-current algebra $\CH_{k'}$.   Then we argue that the
$\CH_{k'}$ currents are gapped out in the flow  leading to the fixed point
$\CG_k /\CH_{k'}$.

\underline{Phase C}  ~~~ Since the only non-zero coupling is
$g_c$ and $g_c \to -\infty$,  this phase clearly has the interpretation
as the fixed point:
\beq
\label{phaseC}
{\rm Phase~ C~ IR~  fixed~ point:} ~~~~~~~~
\frac{osp(4|4)_1}{osp(2|2)_{-2}} 
\eeq

\bigskip
\underline{Phases $S^\pm$ }.   
In  the phases $S^\pm$  the only couplings flowing to $\infty$
are $g_\o , g_s$. 
It was shown in Ref. \cite{BLsqhe} that the operators
$\CO^\o , \CO^s$ are a logarithmic pair.   
One can see explicitly 
from the asymptotics that $g_\o/g_s$ flows  very rapidly
to zero.
We can thus interpret this as $g_\o$ effectively being
zero in comparison to $g_s$.  Since $g_s $ is flowing to $\pm \infty$, 
we propose the fixed point is the coset $osp(4|4)_1/su(2)_0$.

\bigskip
\underline{Phases $SC^\pm$ }.
In this phase $g_c$ decouples in the RG equations and flows to 
$-\infty$ with the same asymptotic as phase $C$.  
Hence both subalgebras $su(2)_0$ and
$osp(2|2)_{-2}$ are massive, and we propose the fixed point
$osp(4|4)_1/su(2)_0\otimes osp(2|2)_{-2} = \emptyset$. 
Since this theory is empty, this is a massive phase. 
This leads to $\Gamma_E = 0$ and a constant density of states
as $E \to 0$.

\def\betaf{{$\beta{\rm eta}$}\ }
\section{Multifractal exponents  from
conformal cosets}

Critical 
exponents related to the multi-fractality
of the wave functions have been the subject of 
numerous 
studies\cite{numerical2,numerical,Zirn1,Zirn2,numerical3}.  
The multi-fractality itself is an important constraint 
on the critical theory.  
These exponents and the correlation length exponent 
$\xi \propto E^{-\nu}$  where $\nu \approx 2.35$\cite{Huck} 
are both important benchmarks for the 
correct critical theory. 
The  multi-fractal exponents are more easily related to 
the conformal scaling dimensions of field operators in the
critical theory\cite{Ludwig}\cite{Mudry}.

In this section we show that the critical theories considered above
lead to multifractality in a very natural way. 
For the original $u(1)$ network model we will consider the 
$osp(2N|2N)_1/u(1)_0$ conformal field theory defined in section IIC,
with stress tensor
\beq
\label{u1stress}
T= T_{osp(2N|2N)_1} - T_{u(1)_0} 
\eeq 
For the spin network model we will consider the  coset
$osp(4N|4N)_1/su(2)_0$.

\def\psib{{\bar{\psi}}}
\def\betab{{\bar{\beta}}}

\def\bPhi{{\bf \Phi}}

For any field $\bPhi (x)$ in an effective field theory
of disorder we can define the quantity
\beq
\label{2.1m}
P^{(q)} = \frac{  \int d^2 x \bar{\langle \bPhi (x) \rangle^q } }
       {\(  \int d^2 x \bar{\langle \bPhi (x) \rangle } \)^q }
\eeq
where $\bar{\langle \bPhi (x) \rangle}$ denotes the disorder
average of the correlation function $\langle \bPhi (x) \rangle$,
and $\bar{ \langle \bPhi (x) \rangle^q }$ denotes the disorder
average of its $q$-th moment in the effective theory
with $N\geq q$ copies.   
The above scales as 
\beq
\label{2.2m} 
P^{(q)} \sim  L^{-\tau (q)}
\eeq
where $L$ is a large length scale.  From simple dimensional
analysis one finds
\beq
\label{2.3m}
\tau (q) = \Gamma(q) - q \Gamma (1) + 2 (q-1) 
\eeq
where $\Gamma(q)$ is the scaling dimension of   
$\bar{\langle \bPhi \rangle^q }  $.       

If $\tau (q)$ is quadratic in $q$,  then imposing $\tau (0) = -2$
and $\tau (1)$ = 1 leads to a one-parameter family
\beq
\label{2.4m}
\tau(q) = (2-\alpha_0) q^2 + \alpha_0 q -2 
\eeq
It is known that usually $\tau (q)$ changes form for
$q$ above some threshold\cite{termination}.  However 
we are interested in the parameter $\alpha_0$  which
is determined near $q=0$ since $\alpha_0 = d\tau/ dq |_{q=0}$,
so we do not consider the phenomenon of multi-fractality
termination.   It is customary 
to define the Legendre 
transform with $\alpha = d\tau / dq$, which defines $q(\alpha)$, 
and 
\beq
\label{2.5m}
f(\alpha ) = \alpha q - \tau (q) = - \frac{(\alpha - \alpha_0)^2}
{4(\alpha_0 -2)} + 2 
\eeq
The fact that the numerical studies show  a close fit to 
the above parabolic form indicates that $\tau(q)$ is 
indeed quadratic in $q$ for small $q$.   
The exponent $\alpha_0$ determines the typical amplitude 
\beq
\label{2.6m}
P_{\rm typical} = \exp \( \bar{ \log \langle \bPhi \rangle } \) 
\sim L^{-\alpha_0} 
\eeq

The multi-fractal dimensions $\Gamma (q)$ follow simply
from the conformal dimensions with respect to the current
algebras using eq. (\ref{dims}) 
where 
$\CG = osp(2N|2N)_1 $ and $\CH = u(1)_0$ or $su(2)_0$,
and $\Delta^{\CG, \CH}$ are conformal scaling dimensions which
follow from the operator product expansion with the stress tensor.

It is convenient to bosonize the fields, where for each copy
\beq
\label{2.8m}
\psi_\pm = e^{\pm i \vphi} , ~~~~~~~~~~~
\beta_+ = e^{i\vphi '} \eta , ~~~~\beta_- = e^{-i\vphi'} \d_z \xi
\eeq
where $\vphi , \vphi'$ are the $z$-dependent parts of
scalar fields $\phi = \vphi (z) + \bar{\vphi}  (\zbar)$,
and $\phi' = \vphi' (z) + \bar{\vphi}' (\zbar)$
and $(\eta, \xi)$ is a conformal dimension $(1,0)$ fermionic
system with central charge $c=-2$\cite{FMS}.
The bosonized action is then
\beq
\label{2.9m}
S_{\rm free}  = \int \frac{d^2x}{8\pi}
\(
\sum_{i=1}^N ~  \d \phi_i \d \phi_i -
\d \phi'_i \d \phi'_i
+ \eta^i \d_\zbar \xi^i + \bar{\eta}^i  \d_z \bar{\xi}^i
\)
\eeq
The $u(1)$ currents are
\beq
\label{2.10m}
H_\psi = \sum_{i=1}^N \psi^i_+ \psi^i_- = i \sum_i \d_z \phi_i
, ~~~~~~
H_\beta = \sum_{i=1}^N \beta^i_+ \beta^i_- = i \sum_i \d_z \phi'_i
\eeq
and similarly for $\bar{H}$.

Consider first the fields $\bPhi = \exp({ia \phi})$ in the  
$u(1)$ theory   labeled by
a parameter $a$.  $\Gamma (q)$ is the dimension of 
\beq
\label{2.12m}
[\bPhi]_q \equiv \bPhi^1 (x) \bPhi^2 (x) \cdots \bPhi^q (x) 
\eeq
where $\bPhi^i$ is the $i$-th copy of the at least $N=q$ 
copy effective theory.   The operator $[\bPhi]_q$ has
dimension $qa^2$ with respect to the $osp(2N|2N)_1$ since
each copy has dimension $a^2$.   
Since  the  $H$  $u(1)$ charge is $aq$,  
eq. (\ref{u1stress})  implies 
\beq
\label{2.13m}
\Gamma (q) = a^2 q (1-q) , ~~~~~\Rightarrow ~~~\alpha_0 = 2+ a^2 , 
\eeq
for $\bPhi = \exp ( i a \phi )$.  

Note that it is important
for the denominator of the coset to be independent of $N$ in
order to obtain this result.  This is equivalent to the statement
that $\Gamma(q)$ can be computed at arbitrary $N\geq q$ 
and should give the same result independent of $N$.  
In particular the $c=-2$ theories $osp(2N|2N)_1/u(1)\otimes u(1)$ 
where the $u(1)$'s are generated by $H_\psi$, $H_\beta$ 
do not have multifractal properties due to the $N$-dependence 
in eq. (\ref{2.24}).

The multi-fractal exponents for the wavefunctions $\Psi (x)$
correspond to choosing $\bPhi$ as the density operator
\beq
\label{2.14m}
\rho   = \psib_- \psi_+ + \psi_- \psib_+ = \cos \phi 
\eeq
The reason is that $|\Psi|^2 \propto G_{\rm ret} (x,x) 
\propto \langle \rho \rangle$.  
Since $\Gamma(1)=0$, this implies a constant density of
states at zero energy.  
Since $\rho $ corresponds to $a=1$, this naively gives 
$\alpha_0 = 3$.   
However one should perhaps bear in mind that the numerical
simulations of the network model used periodic boundary
conditions on the wavefunctions in a cylindrical geometry. 
It is well-known that periodic boundary conditions on the
cylinder are mapped to anti-periodic boundary conditions on 
the plane.   The boundary  conditions of free fermion fields are
modified in the presence of spin, or twist,  fields, which here
are $\sigma_\pm = \exp ({\pm i\phi/2}) $.   This can be seen
from the OPE 
\beq
\label{2.15m}
\psi_\pm (z) e^{\mp i \vphi (0)/2}  \sim \inv{\sqrt{z}} 
e^{\pm i \vphi (0) /2 } 
\eeq
which shows that $\psi_\pm$ picks up a phase $-1$ as $z$ encircles
the origin on the plane.   Since the more relevant operator
in the  OPE of $\rho$ with 
the twist  fields  is again a twist  field,  we suggest that 
$\alpha_0$ may  correspond to $a=1/2$, which gives
$\alpha_0 = 9/4$.    This is quite close to the  
published simulations on the network model\cite{numerical2,numerical3}, 
which report $\alpha_0 = 2.26 \pm .01$ and $2.260\pm .003$. 
The small errors in the latter measurement seem to rule out $9/4$.  
Simulations of other models believed to be in the same
universality class  give the slightly higher result 
$\alpha_0 = 2.28 \pm .03$.   
 
The point-contact conductance is another multi-fractal 
exponent that has been studied numerically for the 
network model\cite{Zirn1,Zirn2}.   Define 
higher moments of the density-density correlation  as 
\beq
\label{2.16m}
G^{(q)} = \bar{  \langle \rho (r) \rho (0) \rangle^q } 
\sim r^{-\tau_G (q)} 
\eeq
where $\rho$ is the density operator.   
Let us assume that  in the flow to the IR there is no mixing between the
various copies.  
Simple dimensional
analysis then gives 
\beq
\label{taug}
\tau_G (q) = 2 \Gamma (q) 
\eeq
where $\Gamma$ is defined the same way as above, i.e. as the
dimension of $\bar{\langle \bPhi \rangle^q}$.  

It is possible to derive a general scaling relation between $\alpha_0$ and
the typical point contact conductance exponent $X_t$.  
Let us generally suppose that
 $\Gamma (q) = Aq + Bq^2$ for some constants $A,B$.  
If the density of states is constant, this implies 
$\Gamma (1) = 0$, or $A=-B$.  Then, for the
point-contact conductance, eq. (\ref{taug}) leads
to an $f(\alpha)$ spectrum 
\beq
\label{2.17m}
f_G (\alpha) = - \frac{ (\alpha - X_t )^2}{4 X_t}  
\eeq
where $X_t = 2A$ is the typical point-contact conductance
$$\exp \( \bar{ \log G} \) \sim r^{-X_t}$$ 
Comparing with eqs. (\ref{2.3m},\ref{2.4m}) we have $\alpha_0 = 2+A$,
which implies the scaling relation:
\beq
\label{2.18m}
X_t = 2 (\alpha_0 -2)
\eeq
We expect the above relation to hold whenever the
density of states is constant\cite{refer}. 
The above relation
works reasonably well using the most recent values of
$X_t$ of $.54 , .57$ reported in \cite{Zirn2} and
the numerical values of $\alpha_0$\cite{numerical,numerical2,numerical3}.  
For $\alpha_0 = 9/4$, one has $X_t = 1/2$.   

Let us turn next to the  coset 
$osp(2N|2N)_1/su(2)_0$.  
The fermion fields transform
as  spin $j=1/2$  doublets under the $su(2)$.  Consider
again the $su(2)$ invariant density operator 
$\bPhi =\rho =  \bar{\psi} \psi$, with dimension $1$ in the 
$osp(4|4)_1$ theory.   Since $su(2)$ primary fields of 
spin $j$ have dimension $\Delta^{su(2)}_j = j(j+1)/2$ 
when $k=0$\cite{KZ},  one finds $\Gamma (1) = 1/4$.  
Since $[\bPhi]_q$ has spin $j=q/2$,  we thus find 
\beq
\label{2.19m}
\Gamma(q) = q - q(q+2)/4    
\eeq
This agrees with ref. \cite{Mudry}  at
small $q$.   
Interestingly, using eqs. (\ref{2.3m},\ref{2.4m},\ref{2.5m}),
the above $\Gamma(q)$ also leads to $\alpha_0 = 9/4$.  
For the point-contact conductance, here the density of
states is not constant, i.e. $\Gamma (1) \neq 0$, so 
eq. (\ref{2.18m})  does not apply.   Rather, 
eq. (\ref{taug})  leads to $X_t = 1$.

If we adopt the viewpoint that the above computations
are a positive indication,  it is tempting to try and
understand the correlation length exponent $\nu$ for
the quantum Hall transition.   This is less straightforward 
since this exponent is not associated with the density operator.
Rather, it is expected to be related to a perturbation of
the action $\delta S = \delta \int d^2 x ~ \Phi_\delta$ 
where tuning $\delta$ through zero corresponds to tuning the
network model to its critical probability.   
As described in section IVA,  this leads to a 
 diverging length 
$\xi \propto \delta^{-\nu}$, where $\nu = 1/(2-\Gamma_\delta)$.  

The challenge is to determine $\Phi_\delta$ from first
principles and this is beyond the scope of this paper. 
Even for the case of the spin quantum Hall transition 
the arguments leading to 
the identification of the  
operator $\Phi_\delta$ which 
gives the percolation exponent $\nu = 4/3$ in this approach
are missing, though there does exist an operator of the right
dimension in the coset 
$osp(4|4)_1 / su(2)_0$\cite{BLsqhe}. (See section IVA.)  
In the percolation picture, $\Phi_\delta$ is the two-hull operator. 
In the remainder of this section  we just explore the 
kinds of exponents that can arise.   

Since the 1-copy theory has a non-critical density of states 
we must consider at least the 2-copy theory.   
Since the density operator is $cos\phi$,
it  is natural to consider
\beq
\label{last}
\Phi_\delta = e^{i\phi_1} e^{- i\phi_2} \> \Phi^1_{\eta/\xi} 
\Phi^2_{\eta/\xi}
\eeq
where $\Phi^i_{\eta/\xi}$ is a field in the i-th $\eta/\xi$ 
copy.   Each $\eta/\xi$ copy describes dense polymers with
a variety of non-trivial dimensions\cite{Saleur}, and dividing
by $u(1)_0$  does not affect these scaling dimensions.  
Interestingly,  there have been some recent attempts at
describing the Quantum Hall transition starting from polymers
rather than percolation\cite{Moore}.  See also ref. 
\cite{Gurarie2,Tsvelik2}. 

Dividing by $u(1)_0$   endows the combined $\exp i\phi$ factors with
dimension $2$.  
If we simply take $\Phi^1_{\eta/\xi}$ as the 1-leg operator
with $\Gamma = -3/16$ and $\Phi^2_{\eta/\xi}$ as a 0-leg, 
or twist, operator with $\Gamma=-1/4$, 
this gives $\Gamma_\delta = 25/16$.   This leads to 
$\nu = 16/7 \approx 2.29$.  
This appears to be the possibility that is
closest to the numerical value of $2.35$  in this scheme\cite{Lec}.   

\section{Conclusions}

We have described how the RG flows based on the all-orders $\beta$ function
for both the Chalker-Coddington and spin network models are attracted to a true
singularity after a finite scale transformation.  The nature of these singularities
remains to be understood.    The RG flows in other domains
are regular and can be continued to arbitrarily large length scales.  

Using an argument that is independent of the $\beta$ function, 
we proposed that the fixed points of general current-current interactions
correspond to solutions of the Virasoro master equation.   For the network
models these correspond to coset conformal field theories, the most promising
being $osp(2N|2N)_1/u(1)_0 $ for the $u(1)$ network and $osp(4N|4N)_1/su(2)_0$ 
for the spin network.  Both of these theories lead to multifractal properties
that we have analyzed.  

For the spin Quantum Hall transition,  it remains an open question    
whether the $osp(4N|4N)_1/su(2)_0$  for $N=1$ is equivalent to percolation.
This can perhaps be answered by comparing with the recent partition functions
in \cite{saleurread}.   It would also be interesting to perform numerical
simulations of the multifractal properties of the spin network model  
and compare them with the coset prediction of $\alpha_0 = 9/4$,  since
this result relies on averages of all higher moments and is thus  not
accessible from percolation.

\section{Acknowledgments}

We would like to thank F. Evers, M. Halpern and D. Khmelnitskii, A. Ludwig,
A. Mirlin  and 
N. Read  for discussions. 
A.L. would like to thank the group at LPTHE, Jussieu, Paris for
their hospitality.  
This work is in part supported by the NSF and the CNRS.

\section{Pauli matrix and intertwiner identities}
 
Using the relations $[\sigma^a, \sigma^b ] = f^{abc} \sigma^c $,
$\{ \sigma^a , \sigma^b \} = \inv{2} \delta^{ab}$ and
$\ep^2= -$1, $(\sigma^a)^\dagger = \sigma^a$ ,
and $\ep \sigma^c = - \sigma^c \ep$,
one can establish the following identity
\beq
\label{identi}
\sigma^a_{ij} \sigma^a_{nm} = \inv{2} \delta_{im}\delta_{jn}
- \inv{4} \delta_{ij} \delta_{nm}
\eeq
The intertwiners satisfy the following identities
\barray
\rhat^a \rho^b = \inv{2} f^{abc} \sigma^c - \inv{2} \delta^{ab}
\quad &,& \quad
\rhat^a \sigma^b = \inv{2} f^{abc} \rhat^c + \inv{4} \delta^{ab}\ep
\non\\
\sigma^a \rho^b = \inv{2} f^{abc} \rho^c + \inv{4} \delta^{ab}\ep
\quad &,& \quad
\rho^a \sigma^b = -\inv{2} f^{abc} \rho^c - \inv{4} \delta^{ab}\ep
\non\\
\rho^a \rhat^b &=& -\inv{2} f^{abc} \sigma^c - \inv{4}\delta^{ab}
\earray


\begin{references}
\bibitem{refer}   After this work was completed, it was pointed out to 
us that this scaling relation  appeared in a footnote in the 
paper\cite{numerical3}.   There, the derivation relies on
a more general result $\tau_G (q) = 2 + \tau(q) + \tau(1-q)$ 
and the form of $\tau(q)$ coming from the theories in \cite{Zirn,Tsvelik0}.
(See section V for definitions.)   Under our assumptions this
is equivalent  to eq. (\ref{taug}).
\end{references}
\end{document}